\documentclass[%
 aip,
 amsmath,amssymb,
 reprint,%
citeautoscript
]{revtex4-2}

\usepackage{filecontents}

\usepackage{graphicx}
\usepackage{dcolumn}
\usepackage{bm}

\usepackage[utf8]{inputenc}
\usepackage[T1]{fontenc}
\usepackage{mathptmx}
\usepackage{etoolbox}

\usepackage[colorlinks,allcolors=black]{hyperref}
\usepackage[capitalise]{cleveref}
\usepackage[acronym]{glossaries}
\usepackage[caption=false]{subfig}
\usepackage{chemformula} 
\usepackage[super]{nth}
\usepackage{physics}
\usepackage[cal=cm]{mathalfa}
\usepackage{siunitx}

\makeatletter
\def\@email#1#2{%
 \endgroup
 \patchcmd{\titleblock@produce}
  {\frontmatter@RRAPformat}
  {\frontmatter@RRAPformat{\produce@RRAP{*#1\href{mailto:#2}{#2}}}\frontmatter@RRAPformat}
  {}{}
}%
\makeatother

\newacronym{md}{MD}{Molecular Dynamics}
\newacronym{qm}{QM}{Quantum Mechanics}
\newacronym{csdf}{CSDF}{Center of the Spin Density Function}
\newacronym{vebe}{VEBE}{Vertical Electron Binding Energy}
\newacronym{aebe}{AEBE}{Adiabatic Electron Binding Energy}
\newacronym{dft}{DFT}{Density Functional Theory}
\newacronym{mp2}{MP2}{Møller--Plesset perturbation theory to \nth{2}-order correction}
\newacronym{ccsdt}{CCSD(T)}{Coupled-Cluster Theory with Singles, Doubles, and noniterative Triples}

\newacronym{oml}{OML}{Orbital Motion-Limited}
\newacronym{rhs}{RHS}{right-hand-side}
\newacronym{lhs}{LHS}{left-hand-side}
\newacronym{lj}{LJ}{Lennard-Jones}
\newacronym{aimd}{AIMD}{Ab Initio Molecular Dynamics}
\newacronym{mb}{MB}{Maxwell-Boltzmann distribution}

\newacronym{si}{SI}{Supporting Information}

\newcommand{\csd}[1]{\rho_s({#1})}
\newcommand{\cvp}{\varepsilon_{\mathrm{0}}}
\newcommand{\micro}{\textmu}
\newcommand{\cdi}{\lambda_{\mathrm{D}i}}

\usepackage{xr}
\makeatletter
\@input{si.aux}
\makeatother
\begin{document}


\title{Can electrostatic stresses affect charged water structures in weakly ionized plasmas?}

\author{Efstratios M. Kritikos}
\affiliation{Department of Applied Physics and Materials Science, California Institute of Technology, Pasadena, 91125, United States}
\email{emk@caltech.edu}
 
\author{William A. Goddard III}
\affiliation{Department of Applied Physics and Materials Science, California Institute of Technology, Pasadena, 91125, United States}

\author{Paul M. Bellan}
\affiliation{Department of Applied Physics and Materials Science, California Institute of Technology, Pasadena, 91125, United States}

\date{\today}

\begin{abstract}
This theoretical and numerical study investigates the impact of electrostatic stresses on the shape of charged water structures (grains) in weakly ionized plasmas. We developed an analytic model to predict the conditions under which a grain in a plasma is deformed. We find that electrostatic stresses can overcome the opposing surface tension stresses on nanometer-scale grains, causing initially spherical clusters to elongate and become ellipsoidal. The exact size limit of the grain for which electrostatic stress will dominate depends on the \textcolor{black}{floating} potential, surface tension, and local radius of curvature. Clusters larger than this limit are not affected by electrostatic stresses due to an insufficient number of electrons on the surface. The model is compared to Molecular Dynamics (MD) simulations performed with a calculated solvated electron potential on initially spherical grains of 2.5 nm radius charged with 0.5 to 1\% electrons. We find excellent agreement between MD simulations and the analytic theory. We also carried out Quantum Mechanics (QM) computations showing that the surface tension increases with decreasing size of the water molecule cluster and increases even more with the addition of solvated electrons. This increase in surface tension can hinder the elongation of the grains. Our QM computations also show that on the nanosecond time scale, the binding force of electrons to water molecule clusters is stronger than the electrostatic repulsion between adjacent electrons, and thus the cluster behaves as an insulator. However, consideration of the very small conductivity of ice shows that on time scales of a fraction of a second, ice clusters behave as conductors, so their surface may be considered to be at an equipotential.
\end{abstract}

\maketitle

\section{Introduction}

Complex or dusty plasmas have astrophysical, meteorological, and technological significance. They are present in interstellar clouds, mesospheric terrestrial clouds, accretion disks, protoplanetary disks, tokamak fusion reactors, and semiconductor processing \cite{Shukla2001,Beckers2023}. \textcolor{black}{Dusty plasmas contain charged solid particles which are known as grains}. The focus of this paper is on charged water grains, but the results are relevant to several other types of dusty plasmas. Grains consisting primarily of water occur in noctilucent clouds \cite{Turco1982}, as part of  Saturn's E-ring \cite{Wahlund2009}, and around the central blue hole of our galaxy \cite{Moultaka2015}. A Caltech experiment has been used to investigate water ice dust grains in a plasma \cite{Nicolov2024}, and we will use results from this experiment to provide relevant parameters for this theoretical investigation.

Grains within a plasma environment accumulate a significant amount of charge. If photoionization and radioactive decay are excluded, this charge originates from the capture of electrons and ions by the grains. Being both hotter and less massive,  electrons have a much higher thermal velocity than ions, causing the grains to become negatively charged. The grain charge plays a pivotal role in grain kinetics and, in certain situations, plasma behavior (e.g., Refs. \citenum{Wardle2007,Okuzumi2009}). Theories of grains in a dusty plasma typically assume that the grains are spherical (e.g., Refs. \citenum{Goertz1989,Shukla2001,Mcclure2012}). However, several studies have shown compelling evidence that grains can be elongated with an approximately ellipsoidal shape \cite{Rapp2007,Chai2018,Bellan2020} or completely irregular \cite{Huang2012,Potapov2021}. Ellipsoidal ice grains have been observed in the Caltech experiment \cite{Marshall2017,Nicolov2024}. 

Grains can depart from a spherical shape as a result of both internal and external factors. Externally, large grains can form due to the collision and aggregation of smaller clusters \cite{Dominik1997,Dominik2006}. However, there is experimental evidence of grain growth into elongated, nonspherical shapes even when grains do not collide \cite{Marshall2017}. Another external factor that can cause the elongated shape of the grain is the preferential accumulation of water molecules at sharp protrusions\cite{Nemchinsky2018,Bellan2020}. Specifically, water molecules, being polar, are attracted by the strong electric field gradient at sharp protusions, causing the grain to grow unevenly. 

This raises the question of how grains acquire a sharp protrusion in the first place. A deviation from sphericity can be created by internal factors. Specifically, there is a competition within the grain between outward-directed electron-electron repulsions and inward-directed surface tension of the material. If the electrostatic stresses surpass the surface tension at some location, then the grain shape changes \cite{Bellan2020}. An analysis of these stresses must also consider the impact of nanosized curvatures and electrons on the surface tension stresses. The interaction between surface tension and electrostatic stresses acting on \textcolor{black}{ellipsoidal grains} has not been previously investigated. Additionally, to the best of the authors' knowledge, the impact of charges on the surface tension of charged water clusters is unknown.

The charge distribution inside the grain affects the resulting internal stresses. If the grains are conducting, then all electrons reside on the surface. On the other hand, the electric potential of non-conducting grains can be arbitrary and can vary considerably throughout the grain volume. Theories predicting the charge of a grain within a plasma assume that grains are conducting (e.g., Refs. \citenum{Shukla2001,Bellan2008}). Nonetheless, the conductivity assumption of water-based grains has not been previously assessed. Such an assumption is crucial since a completely different charging process would be expected if grains were insulating rather than conducting. Finally, to understand the conditions under which grains undergo morphological changes, it is necessary to identify the timescales of the various phenomena occurring in a dusty plasma, such as plasma kinetics, grain charging, grain behaving as a conductor, and thermodynamic equilibrium.

In this study, a comprehensive theoretical investigation of the electrostatic physics and surface tension of water grains inside a weakly ionized plasma is performed. \textcolor{black}{The analysis can be directly applied to other materials.} As part of this investigation, \gls{md} simulations of grains consisting of water molecules and electrons are performed for the first time to study the impact of electrostatic repulsions on the structure of grains. In addition, \gls{qm} computations are used to investigate the effect of electrons on the potential energy and surface tension of small clusters of water molecules.

The objectives of this work are to: (i) determine the conditions under which electrostatic stresses compete with the surface tension stresses in spherical and ellipsoidal shapes; (ii) validate the theory using \gls{md} simulations of water grains (iii) investigate the impact of cluster size and solvated electrons on the surface tension using \gls{qm} computations; (iv) evaluate the validity of the conductivity assumption in the nanoscale using \gls{qm} simulations and in the macroscale using an analytic model.

The structure of the paper is as follows: first, the charging process of grains in a plasma is discussed. Then, the competition between electrostatic and surface tension stresses on grains is modeled. This model is validated using \gls{md} simulations. Finally, the conductivity approximation of water grains is assessed, and the surface tension of nanosized clusters is investigated. In addition, we provide a \gls{si} document, which contains numerical methods, extra calculations and graphs, and simulation details. The \gls{si} is of a technical nature that supports the findings presented in this paper, but incorporating it might detract from the logic of the main paper. References to the \gls{si} are provided within the text, where appropriate, so the interested reader can follow up on the detail.

\section{Floating potential}
\label{sec:net_charge}

\begin{table}[!b]
    \centering
    \begin{tabular}{cc}
    Parameter & Value \\
    \hline
        $m_i$ & 39.948 amu (Argon) \\
        $n_{n}$ & $10^{21}$ m$^{-3}$ \\
        $n_i$ & $10^{15}$ m$^{-3}$ \\
        Fractional ionization & $10^{-6}$ \\
        $n_d$ &  $2\times 10^{11}$ m$^{-3}$ \\
        $T_i$ & 30 meV \\
        $T_e$ & 2 eV \\
    \hline
    \end{tabular}
    \caption{Parameters for the computation of dust grain charges.}
    \label{tab:q_d_param}
\end{table}

This section discusses the net charge that the grains acquire and their electron distribution. The net charge of a spherical grain in a plasma is given by the \gls{oml} theory \cite{Bellan2008}. The \gls{oml} theory is accurate when ions and electrons are collisionless within a sphere centered on the grain and having a radius equal to the ion Debye length $\cdi$. In addition, the theory is accurate for grains with sizes significantly smaller than $\cdi$ \cite{Lampe2001}. An important assumption of the \gls{oml} theory is that the grain is conducting and thus its surface is an equipotential. We will make this assumption here and evaluate it in \cref{sec:solv_cond}. According to \gls{oml}, the \textcolor{black}{floating} potential $\varPhi$ that the grains acquire inside a dusty plasma is given by the solution of the equation \cite{Shukla2001}
\begin{equation}
    \dfrac{1}{\psi}-\left(1 + \dfrac{1}{\psi}\right) \sqrt{\dfrac{m_e T_i}{m_i T_e}} \exp(\dfrac{\psi T_i}{T_e})=3\dfrac{r_d \cdi^2}{r_{WS}^3},
    \label{eq:psi_d}
\end{equation}
where the dimensionless potential of the grain is 
\begin{equation}
\psi=-\dfrac{e\varPhi}{\kappa T_i},
\end{equation}
$r_d$ is the radius of the spherical dust grain, $\kappa$ is Boltzmann's constant, and $e$ is the electron charge. $T$ and $m$ are the temperatures and masses of the ions (subscript $i$) or electrons (subscript $e$). The Wigner-Seitz radius $r_{WS}$ is the nominal spacing between adjacent grains and is computed from the empty volume of the system. Finally, the ion Debye length is 
\begin{equation}
\cdi=\sqrt{\dfrac{\cvp \kappa T_i}{n_i e^2}},
\label{eq:lambda_di}
\end{equation}
where $\cvp$ is the vacuum permittivity and $n_i$ is the volume-averaged ion density.

The charge $Q$ of the grain is computed from
\begin{equation}
    Q=4 \pi \cvp r_d \varPhi =
    4 \pi e \cdi^2 n_i r_d \psi.
    \label{eq:q}
\end{equation}

\begin{figure}[!tb]
    \centering
    \includegraphics[width=0.5\linewidth]{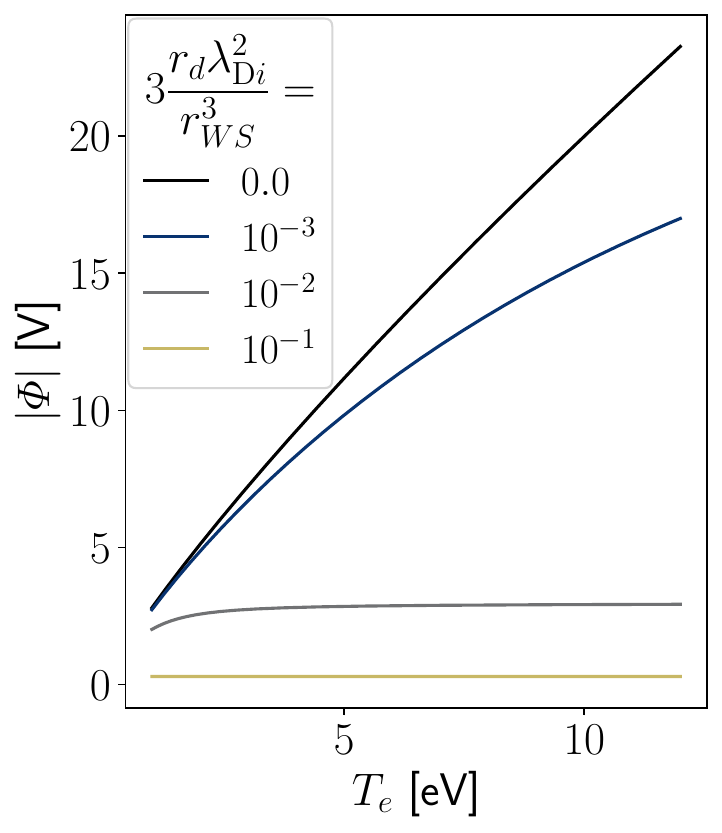}%
    \includegraphics[width=0.5\linewidth]{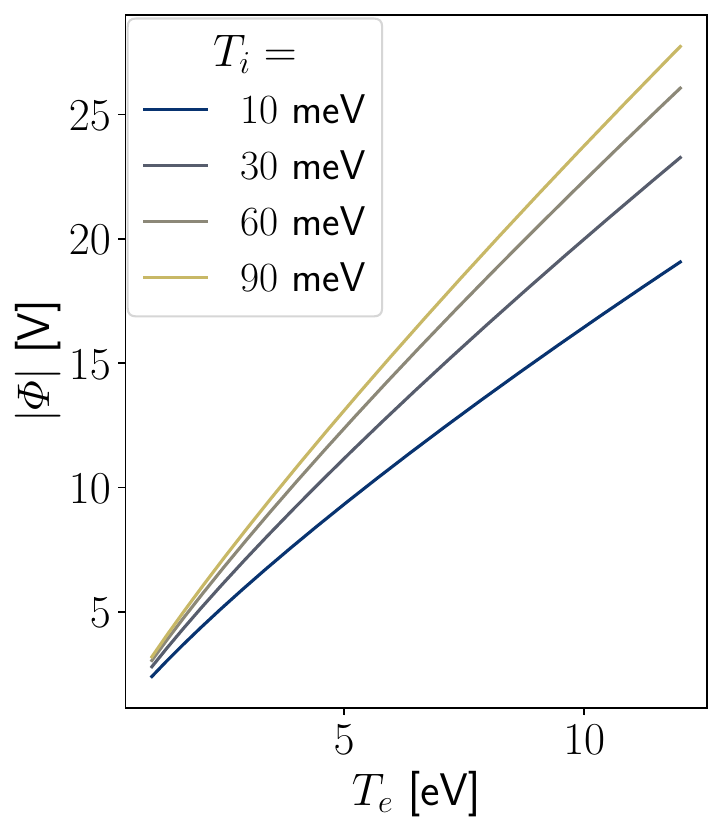}
    \caption{Parametric evaluation of \cref{eq:psi_d}, under different (a) values of the \gls{rhs} at $T_i=30$ meV and (b) $T_i$ values at $3r_d \cdi^2/r_{WS}^3=0$.}
    \label{fig:Phi_parametric}
\end{figure}

The nominal parameters of the Caltech dusty plasma experiment \cite{Nicolov2024} are listed in \cref{tab:q_d_param}. These parameters are considered to be similar to the parameters of various astrophysical situations, but should also be relevant to certain terrestrial and industrial situations.  The parameters assumed in the analysis presented in this paper are based on \cref{tab:q_d_param} unless otherwise specified. \Cref{fig:Phi_parametric} shows the \textcolor{black}{floating} potential of a spherical grain as a function of $T_e$ for four different values of the \gls{rhs} of \cref{eq:psi_d} and four different values of $T_i$. This dependence describes the following behavior: Increasing the ion or electron temperatures increases the \textcolor{black}{floating} potential of the grains. When the dust grains are large and densely packed, i.e., large \gls{rhs} of \cref{eq:psi_d}, they do not individually acquire much charge. On the contrary, small and sparse grains, i.e., small \gls{rhs}, can \textcolor{black}{attain} a \textcolor{black}{floating} potential \textcolor{black}{that is} more than twice the electron temperature. The quantity $3\psi r_d \cdi^2/r_{WS}^3$ is the fraction of all the electrons that have been captured by the dust grains. \Cref{fig:Phi_parametric}(left) shows that the dust \textcolor{black}{grain floating} potential is higher when most of the electrons are free rather than captured by dust. It is noted that the charging time of grains, $\tau_{q}$, is inversely proportional to the grain size and the ion density, i.e., $\tau_{q}\propto r_d^{-1}n_i^{-1}$ \cite{Shukla2001,Merlino2021}. Therefore, grains get charged more slowly when they are smaller or within a lower-density plasma. A theoretical lower limit for $\tau_q$ is derived in \cref{sec:md_valid}.

The floating potential differs for non-spherical grains. To the authors' knowledge, an \gls{oml} formulation for ellipsoidal grains has not been \textcolor{black}{reported in the literature}. Below, we derive the  \gls{oml} floating potential for cylindrical grains, which can be considered an approximation for highly prolate ellipsoidal grains. The end effects of the cylindrical grains are neglected; thus, the collection of charged species occurs only on the side surface of the cylinder. The \textcolor{black}{ion} and \textcolor{black}{electron} currents \textcolor{black}{flowing from the surrounding plasma onto a} cylinder of radius $r_d$ and length $l$ are given by Allen \cite{Allen1992} \textcolor{black}{as}
\begin{equation}
    I_{i}=2 e \pi n_i r_d l \sqrt{\dfrac{\kappa T_i}{2\pi m_i}}\left[\dfrac{2\sqrt{\psi}}{\sqrt{\pi}}+e^\psi \mathrm{erfc}\left(\sqrt{\psi}\right)\right],
    \label{eq:curr_i_cyl}
\end{equation}
and 
\begin{equation}
    I_{e}=-e n_i r_d l \sqrt{\dfrac{2\pi \kappa T_e}{m_e}} \exp(-\psi),
    \label{eq:curr_e_cyl}
\end{equation}
where the complementary error function is $\mathrm{erfc}(x)=1-\mathrm{erf}(x)$, and $\mathrm{erf}(x)=(2/\sqrt{\pi})\int_0^x \exp(-t^2)dt$. Following Allen \cite{Allen1992}, the term inside the brackets in \cref{eq:curr_i_cyl} can be approximated to $2\sqrt{\psi}/\sqrt{\pi}+e^\psi \mathrm{erfc}(\sqrt{\psi})\approx 2\sqrt{(1+\eta)/\pi}$ for $\psi>2$.

\begin{figure}[!tb]
    \centering
    \includegraphics[width=0.8\linewidth]{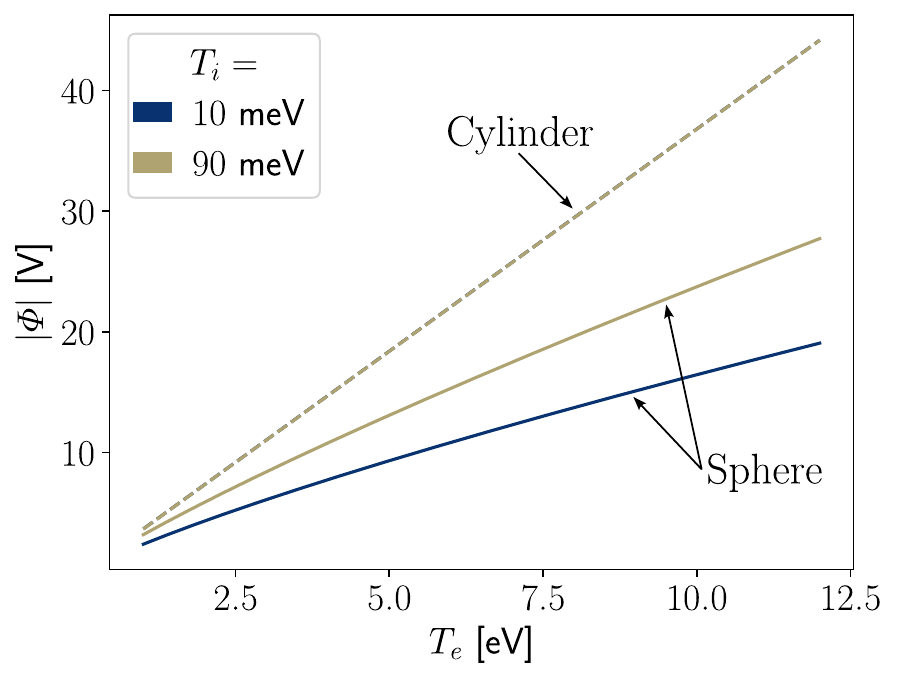}
    \caption{Comparison of $\varPhi$ obtained from \cref{eq:psi_d} for spherical grains (solid line) and from \cref{eq:psi_d_cyl} for cylindrical grains (dashed line) under different $T_i$ values at $3r_d \cdi^2/r_{WS}^3=0$.}
    \label{fig:Phi_parametric_cyl}
\end{figure}

Based on these currents, the floating potential can be determined for cylindrical grains. Specifically, considering that $I_i+I_e=0$ and neglecting electron depletion (i.e., the term $3 r_d \cdi^2/r_{WS}^3$ in \cref{eq:psi_d}), the \gls{oml} theory of cylindrical grains becomes
\begin{equation}
    1-2\sqrt{1+\psi} \sqrt{\dfrac{m_e T_i}{\pi m_iT_e}} \exp(\dfrac{\psi T_i}{T_e}) = 0.
    \label{eq:psi_d_cyl}
\end{equation}
A comparison of the floating potentials, obtained from \cref{eq:psi_d,eq:psi_d_cyl} for spherical and cylindrical grains, respectively, is shown in \cref{fig:Phi_parametric_cyl}. \textcolor{black}{The floating potential of either spherical or cylindrical grains in \cref{fig:Phi_parametric_cyl} does not depend on the grain dimensions since electron depletion has been neglected.} It is seen that the floating potential of a cylindrical grain is significantly larger than that of a spherical grain. In addition, the floating potential of cylindrical grains has negligible dependence on ion temperature. It is expected that the \textcolor{black}{floating} potential of an ellipsoidal grain will lie in between the two extreme cases of spherical and cylindrical grains. Thus, the floating potential of an ellipsoidal grain is expected to increase with increasing aspect ratio.

\textcolor{black}{It is conceivable that electron-electron repulsions could limit the net charge of very small grains; the associated electron expulsion is called Electron Field Emission (EFE). Specifically, strong electrostatic repulsions could force electrons to tunnel through the potential energy barrier of the grain surface and escape. In \cref{sec:solv_cond} we investigate how the binding energy of solvated electrons in water clusters competes with these electrostatic repulsions. We find that electrostatic repulsions become insignificant above interelectron distances of 3 {\AA} for the bulk material and as low as 8 {\AA} for small water molecule clusters.}

\textcolor{black}{The standard analytical EFE formula (e.g., Refs. \citenum{Stefanovic2006,Vaverka2014,Kyritsakis2015,Holgate2017a,Holgate2018}) is based on a free-electron assumption and so only applies to metallic surfaces. In \cref{sec:si_solv_cond} we compare the predictions of the standard analytical EFE formula for electrically isolated metallic spheres\cite{Holgate2018} to our \gls{qm} simulations. Due to the free-electron approximation (i.e., metal assumption), the standard analytical EFE formula predicts significant electron-electron repulsions, whereas the first-principle \gls{qm} simulation presented here predicts negligible repulsions. We conclude that the standard EFE formula significantly overestimates EFE for the grains investigated here and thus is not applicable. Future research should be performed to develop an EFE prediction relevant to non-metallic spherical and ellipsoidal grains.}

\textcolor{black}{If situations develop where EFE becomes significant so mutual electron repulsion constrains the number of electrons on small grains, these grains may still develop significant net negative charge from accumulation of anions such as \ch{OH-} or \ch{O-} coming from the plasma\cite{Bellan2022} instead of electrons. Because anions are heavier than electrons, they are less likely to escape. Grains composed of both solvated electrons and anions should be investigated in the future.}

\section{Electron distribution}
\label{sec:el_dist}

\begin{figure}[!htb]
    \centering
    \includegraphics[width=0.65\linewidth]{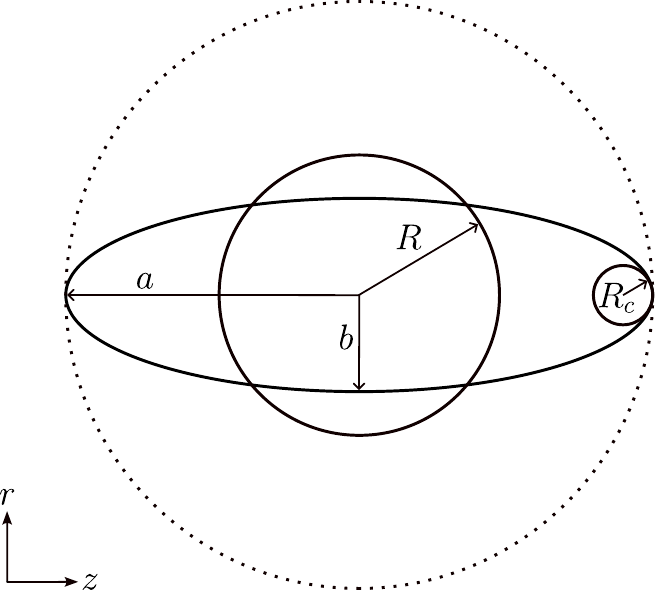}
    \caption{Characteristic lengths and shapes related to  an ellipsoid: (i) the ellipsoid itself (solid line) with major radius $a$ and minor radius $b$ (ii) a same-volume sphere (solid line with  radius $R$), (iii) radius $R_c$ osculating circle tangent to  the tip of the ellipsoid, (iv) the smallest sphere enclosing the ellipsoid (dotted line with radius $a$)}
    \label{fig:ellipse_sphere_lengths}
\end{figure}

We continue the analysis by focusing on the electrodynamics of ellipsoidal grains. The surface of an ellipsoid with a major radius $a$ and two equal minor radii $b$ is prescribed in Cartesian and cylindrical coordinates by
\begin{equation}
    \dfrac{x^2+y^2}{b^2}+\frac{z^2}{a^2}=\dfrac{r^2}{b^2} + \dfrac{z^2}{a^2} = 1.
    \label{eq:ell_coord}
\end{equation}
Following Ref.~\citenum{Bellan2020}, we define the aspect ratio $\alpha = a/b$. Thus, a prolate ellipsoid has $\alpha > 1$. Also, we define $R$ as the radius of a sphere having the same volume as the ellipsoid, i.e., $4\pi ab^2/3=4\pi R^3/3$. Therefore, the major and minor radii of the ellipsoid are related to $R$ by
\begin{equation}
    a=\alpha^{2/3} R,
    \label{eq:ell_a}
\end{equation}
\begin{equation}
    b=\alpha^{-1/3} R.
    \label{eq:ell_b}
\end{equation}
\Cref{fig:ellipse_sphere_lengths} shows the various lengths associated with an ellipsoid and a same-volume sphere. This figure also indicates $R_c$, the radius of curvature of the sharp tip (vertex) of the ellipsoid and the smallest sphere enclosing the ellipsoid (dotted line); the relevance of $R_c$ and the enclosing sphere will be discussed in later paragraphs.

\Cref{eq:ell_coord} can be expressed in terms of $R$ and $\alpha$ as
\begin{equation}
\begin{aligned}
    r(z)&=\alpha^{-1/3}\sqrt{R^{2}-\alpha^{-4/3} \, z^{2}}.
    \label{eq:ell_coord_rx}
\end{aligned}
\end{equation}

The capacitance of a perfectly conducting ellipsoid is~\cite{Shumpert1972,Bellan2020}
\begin{equation}
    C = \dfrac{8 \pi \cvp \alpha^{2/3} R \sqrt{1-\alpha ^{-2}}}{\ln{\left(\frac{1+\sqrt{1-\alpha ^{-2}}}{1-\sqrt{1-\alpha ^{-2}}}\right)}}.
    \label{eq:c_ell}
\end{equation}

\begin{figure}[!b]
    \centering
    \includegraphics[width=0.8\linewidth]{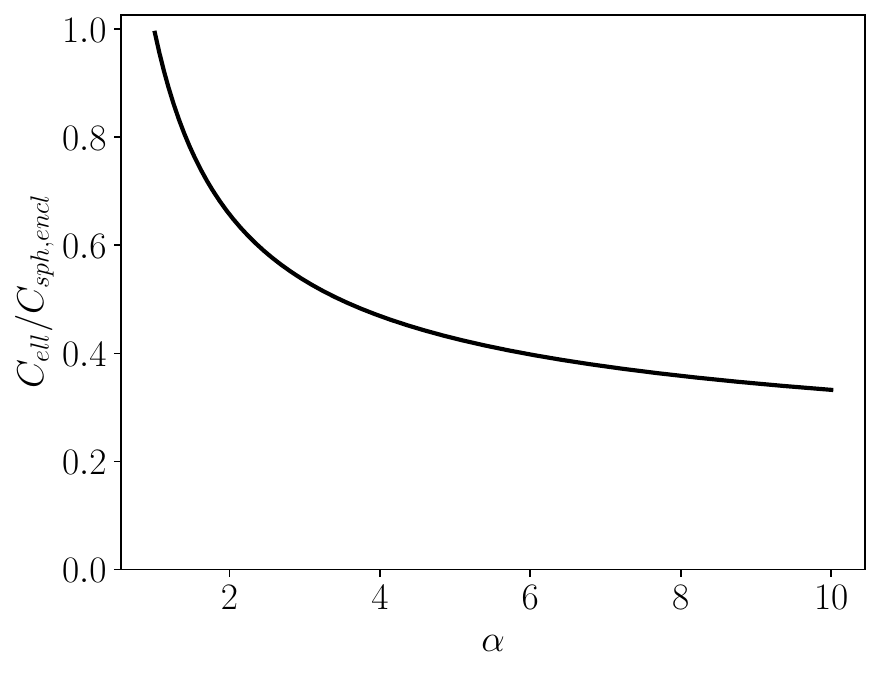}
    \caption{Plot of $C_{ell}/C_{sph,encl}$, i.e., the ratio of the capacitance of an ellipsoid to the capacitance of the smallest enclosing sphere.}
    \label{fig:capacitance_sphere_ellipsoid}
\end{figure}

It is instructive to compare the capacitance of a large aspect ratio ellipsoid to that of the smallest sphere that would enclose the ellipsoid, that is, the capacitance of an ellipsoid having $a \gg b$ compared to the dotted line sphere in \cref{fig:ellipse_sphere_lengths}. The capacitance of this radius $a$ sphere is $C_{sph,encl}=4 \pi \cvp a$. For large aspect ratios, the ratio of the two capacitances reduces to $C_{ell}/C_{sph,encl} \approx 2/ (1.38 +2 \ln\alpha)$, which is of order unity.  \Cref{fig:capacitance_sphere_ellipsoid}  plots $C_{ell}/C_{sph,encl}$ as a function of aspect ratio. An ellipsoid will thus hold approximately the same amount of charge as the smallest enclosing sphere (dotted line in \Cref{fig:ellipse_sphere_lengths}). This conclusion agrees with experimental and \gls{md} studies\cite{Asnaz2018,VanMinderhout2021}, which show that the charge of an irregularly shaped particle within a plasma is of the order of the charge of the smallest enclosing spherical particle when both are charged to the same potential. This conclusion means that the charge to mass ratio $Q/m$ of a spherical grain charged to some given potential will increase when the grain is deformed into a same-volume ellipsoid because the charge $Q$ will increase, whereas the mass $m$ will stay the same. \Cref{eq:c_ell} shows that if $1<\alpha <10$, the $Q/m$ of an ellipsoid compared to the $Q/m$ of a same-volume sphere scales approximately as $1+0.062(\alpha-1)$.  Thus, for example, an ellipsoid with $\alpha=10$ has an approximately 50\% larger $Q/m$ ratio than the same-volume sphere charged to the same potential.

The \textcolor{black}{surface} charge density of a conducting ellipsoid with a total charge of $Q$ is \cite{Smythe1988}
\begin{equation}
\sigma(r,z)=\frac{Q}{4\pi b^2a\sqrt{\frac{r^2}{b^4}+\frac{z^2}{a^4}}},
\label{eq:sigma_ell}
\end{equation}
where $r$ and $z$ are respectively the axes parallel to the minor and major radii of the grain. Using \cref{eq:ell_coord_rx,eq:ell_a,eq:ell_b}, the surface charge density is given with respect to $z$ as
\begin{equation}
\begin{aligned}
\sigma(z)&=\frac{Q}{4\pi \alpha^{1/3}R\sqrt{R^2-\frac{z^2}{\alpha^{4/3}}\left(1-\frac{1}{\alpha^2}\right)}}
\label{eq:sigma_ell_2}.
\end{aligned}
\end{equation}
We now define $dQ_{z_1,z_2}$ to be the charge on the grain surface between axial position $z_1$ and axial position $z_2$. By integrating $2\pi\sigma(z)r(z)\dd z$ over the interval $[z_1,z_2]$, $dQ_{z_1,z_2}$ is given by
\begin{equation}
\begin{aligned}
    dQ_{z_1,z_2}&=\dfrac{Q}{2\alpha^{1/3}R^{2}}\int_{z_{1}}^{z_{2}} \dfrac{\alpha^{-1/3}R\sqrt{1-\frac{z^{2}}{\alpha^{4/3}R^{2}}}}{\sqrt{1-\frac{z^{2}}{\alpha^{4/3}R^{2}}\left(1-\frac{1}{\alpha^{2}}\right)}} \dd z,
    \label{eq:z_d_z_ell_full}
\end{aligned}
\end{equation}
where \cref{eq:ell_coord_rx} was used for the $r$ dependence. To avoid solving the integral in \cref{eq:z_d_z_ell_full} analytically, we focus on ellipsoidal grains with $\alpha \gg 1$. In that case, $1-\alpha^{-2}\approx 1$ and thus \cref{eq:z_d_z_ell_full} becomes
\begin{equation}
\begin{aligned}
   dQ_{z_1,z_2}&\approx \dfrac{q}{2\alpha^{2/3}R}\left(z_{2}-z_{1}\right).
    \label{eq:z_d_z_ell}
\end{aligned}
\end{equation}
Therefore, the number of electrons per axial length is independent of axial position for a conducting ellipsoid with a large aspect ratio. \textcolor{black}{It should be noted that an uneven charge density along the surface of the ellipsoidal grain could affect the collection of the orbiting ions and electrons. This should be investigated in the future.}

\begin{figure}[!htb]
    \centering
    \includegraphics[width=0.8\linewidth]{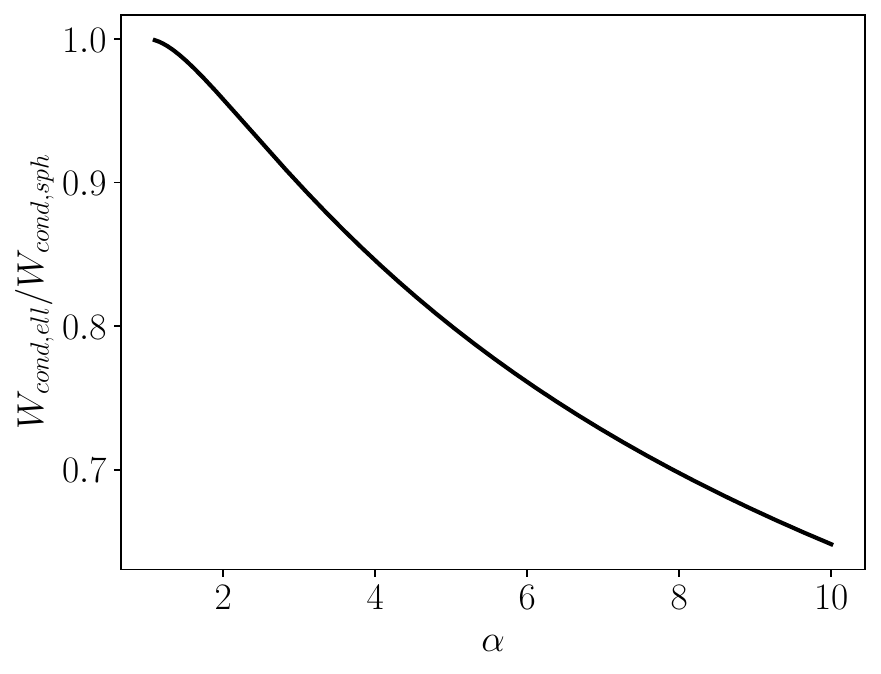}
    \caption{The ratio of the necessary work to construct the charge distribution of a conducting ellipsoid compared to a sphere charged to the same potential as a function of the ellipsoid's aspect ratio.}
    \label{fig:work_ratio}
\end{figure}

We now examine the stability of spherical and ellipsoidal grains by evaluating the work to create a charge distribution~\cite{Griffiths2013}
\begin{equation}
W=\dfrac{1}{2}\int \rho_q \varPhi\, \dd\bm{V},
\end{equation}
where $\rho_q$ is the charge density and $\dd\bm{V}$ is the differential of volume. For a conductor, all the charge lies on the surface, so $\rho_q \varPhi\, \dd\bm{V} \rightarrow \sigma \varPhi_{sfc} dS$, where $\varPhi_{sfc}$ is the surface potential.
For a conducting sphere, $\sigma=Q/(4\pi R^2)$ and $\varPhi=Q/(4\pi \cvp R)$ so
\begin{equation}
W_{cond,\ sph}=\dfrac{Q^2}{8 \pi \cvp R}.
\label{eq:work_cond}
\end{equation}
For a conducting ellipsoid, $\sigma$ is given by \cref{eq:sigma_ell_2}, and the surface potential is given by~\cite{Smythe1988,Bellan2020}
\begin{equation}
    \varPhi=\frac Q{8\pi\cvp \alpha^{2/3} R \sqrt{1-1/\alpha^2}}\ln(\frac{1+\sqrt{1-1/\alpha^2}}{1-\sqrt{1-1/\alpha^2}}).
\end{equation}
Thus, the work required to create the charge distribution of a conducting ellipsoid is
\begin{equation}
\begin{aligned}
    W_{cond,ell} = &\frac{Q^{2}}{16\pi \cvp \alpha^{2/3}R\sqrt{1-1/\alpha^{2}}}\ln(\frac{1+\sqrt{1-1/\alpha^{2}}}{1-\sqrt{1-1/\alpha^{2}}}).
    \label{eq:work_cond_ell}
\end{aligned}
\end{equation}
The ratio of the work required to create an elongated conducting ellipsoid compared to that of a same-volume conducting sphere is plotted in \cref{fig:work_ratio} and is approximately equal to $\ln(4\alpha^2)/(2\alpha^{2/3})$ for $\alpha^2>>1$. Hence, when charged to the same potential, a conducting ellipsoidal grain is in a lower energy state than a spherical one. In addition, its electrostatic energy decreases as it becomes more elongated. Another way of seeing this is to recall that a large aspect ratio ellipsoid has approximately the same capacitance as the smallest enclosing sphere (i.e., a sphere with radius $a$). Since the radius of the smallest enclosing sphere is larger than the radius of a same-volume sphere (i.e., $a > R$) and since capacitance is proportional to radius, the smallest enclosing sphere has more capacitance than a same-volume sphere. Since the energy stored in a capacitor is $W=C\Phi^2/2 = Q^2/2C$, the energy of an ellipsoid is lower than that of a same-volume sphere with the same charge. This means that it is energetically favorable for a sphere to deform into an ellipsoid if the amount of charge is fixed.

\section{Electrostatic and surface tension stresses}
\label{sec:elec_surf_stress}

In this section, the competition between electrostatic stress $\tau_{\mathcal{E}}$ and surface tension stress $\tau_{\gamma}$ is investigated. The grains are approximated as conducting spheres and axially symmetric ellipsoids (prolate spheroids). These are reasonable approximations compared to the grain shape observed experimentally \cite{Chai2018,Nicolov2024} and have been applied in other theoretical studies (e.g., Refs. \citenum{Dolginov1976,Gledhill2000,Rapp2007,Bellan2020}). Surface tension \textcolor{black}{produces} an inward force, whereas repulsions \textcolor{black}{between} electrons \textcolor{black}{produce} an outward force. Therefore, surface tension opposes the tendency of a sphere to become an ellipsoid and instead creates a force that tends to make an ellipsoid become a sphere. The electrostatic stress dominates when the condition
\begin{equation}
\tau_{\mathcal{E}} \ge \tau_{\gamma}
\label{eq:stresses_ineq}
\end{equation}
is satisfied. The analysis is performed at the tip of the ellipsoid, i.e., at $z=a$ and $r=0$ (cf. \cref{fig:ellipse_sphere_lengths}). This \textcolor{black}{location} has the smallest radius of curvature and the maximum stresses. The limiting case of a spherical grain \textcolor{black}{corresponds to} $\alpha \rightarrow 1$. The local stress (force normal to \textcolor{black}{the} surface \textcolor{black}{per unit area}) from surface tension $\gamma$ is \cite{wilson1925,taylor1964}
\begin{equation}
\begin{aligned}
\tau_{\gamma}=\gamma\left(\dfrac{1}{R_1}+\dfrac{1}{R_2}\right),
\label{eq:tau_gamma}
\end{aligned}
\end{equation}
where $R_1$ and $R_2$ are the local orthogonal radii of curvature of the surface. For the ellipsoid under consideration, $R_1=R_2=R_c$, where $R_c$ is indicated in \cref{fig:ellipse_sphere_lengths}. Therefore, $\tau_{\gamma}=2\gamma/R_c$. For water at 0 {\textcelsius}, the surface tension of solid-vapor (SV), liquid-vapor (LV), and solid-liquid (SL) are $\gamma_{SV}=109$ mN/m, $\gamma_{LV}=76$ mN/m, and $\gamma_{SL}=33$ mN/m, respectively.\cite{Hobbs2010} Note that the surface tension for solid-liquid or solid vapor interfaces is known in the literature as specific surface-free energy or surface interaction energy and has units of energy per surface area. Since these two terms are only semantically different, they will both be referred to as surface tension \textcolor{black}{in} the remainder of this paper. The radius of curvature of a curve with functional form $z(r)$ is \cite{love1918}
\begin{equation}
R_c=\left|\dfrac{\left[1+\left(\frac{dz}{dr}\right)^{2}\right]^{3/2}}{\frac{d^2z}{dr^2}}\right|.
\label{eq:radius_curvature}
\end{equation}
From \cref{eq:ell_coord}, $z(r) = a \sqrt{1-r^2/b^2}$  and thus
\begin{equation}
\begin{aligned}
\dfrac{dz}{dr} = -\dfrac{a}{b^2} r \left(1-\dfrac{r^2}{b^2}\right)^{-1/2}
     \label{eq:ell_coord_dzdr}
\end{aligned}
\end{equation}
\textcolor{black}{while}
\begin{equation}
\begin{aligned}
\dfrac{d^2 z}{d r^2} =-\dfrac{a}{b^2}\left(1-\dfrac{r^2}{b^2}\right)^{-3/2}.
\label{eq:ell_coord_d2zdr2}
\end{aligned}
\end{equation}
Upon inserting \cref{eq:ell_coord_dzdr,eq:ell_coord_d2zdr2} into \cref{eq:radius_curvature}, the radius of curvature is found to be
\begin{equation}
R_c= \dfrac{1}{ab} \left[b^2+\left(\dfrac{a^2}{b^2}-1\right)r^2\right]^{3/2}.
\label{eq:radius_curvature_ell}
\end{equation}
At the tip of the ellipsoid $r=0$, so the radius of curvature is 
\begin{equation}
R_c\Big|_{r=0}=\dfrac{b^2}{a}=\dfrac{R}{\alpha^{4/3}}.
\label{eq:radius_curvature_ell_final}
\end{equation}

The electrostatic stress is equal to
\begin{equation}
\tau_{\mathcal{E}}=\dfrac{\cvp \mathcal{E}^2}{2}=\dfrac{\sigma^2}{2\cvp}.
\label{eq:elec_stress}
\end{equation}
\textcolor{black}{Using} \cref{eq:sigma_ell_2}, the charge density \textcolor{black}{at the ellipsoid tip  is given by}
\begin{equation}
\sigma_{cond}(a)=\dfrac{\alpha^{2/3}Q}{4\pi R^{2}}.
\label{eq:sigma_cond}
\end{equation}
Under the assumption that the grain is a conductor, the net charge of the ellipsoid is $Q=C\varPhi$. Therefore, using \cref{eq:c_ell}, the surface charge density at the ellipsoid tip is
\begin{equation}
\begin{aligned}
\sigma_{cond}(a) =\frac{2\cvp\alpha^{4/3}\sqrt{1-1/\alpha^{2}}}{\ln\left(\frac{1+\sqrt{1-1/\alpha^{2}}}{1-\sqrt{1-1/\alpha^{2}}}\right)}\frac{\varPhi}{R}.
\label{eq:sigma_tip}
\end{aligned}
\end{equation}

The surface tension $\gamma$ of a small water cluster depends on the local radius of curvature and is given by Tolman~\cite{Tolman1949} as 
\begin{equation}
\begin{aligned}
\gamma = \dfrac{\gamma_{\infty}}{1+\dfrac{2\delta}{R_c}} = \dfrac{\gamma_{\infty} R}{R+2\alpha^{4/3}\delta}.
\label{eq:gamma_Tolman_curvature}
\end{aligned}
\end{equation}
The parameter $\delta$, called Tolman length, becomes important at very small $R_c$.  A detailed discussion of $\delta$ will be given in \cref{sec:surf_tens}, where, using QM computations, $\delta$ is found to be negative and of the order of 1 {\AA}.

Using \cref{eq:sigma_tip} in \cref{eq:elec_stress} and \cref{eq:gamma_Tolman_curvature} in \cref{eq:tau_gamma} the inequality given by \cref{eq:stresses_ineq} becomes 
\begin{equation}
\begin{aligned}
\underbrace{\dfrac{\left(\alpha^{2}-1\right)\left(R+2\alpha^{4/3}\delta\right)}{\alpha^{2/3}R^2 \left[\ln\left(\frac{1+\sqrt{1-1/\alpha^{2}}}{1-\sqrt{1-1/\alpha^{2}}}\right)\right]^{2}}}_{\mathrm{LHS}} &\ge \underbrace{\dfrac{\gamma_{\infty}}{\cvp \varPhi^{2}} .
\vphantom{\dfrac{1}{\left[\ln\left(\frac{1+\sqrt{1-1/\alpha^{2}}}{1-\sqrt{1-1/\alpha^{2}}}\right)\right]^{2}}} 
}_{\mathrm{RHS}}
\label{eq:est_ineq}
\end{aligned}
\end{equation}

\Cref{eq:est_ineq} can be expressed as a quadratic equation in $R$ with two roots. The roots are
\begin{widetext}
\begin{equation}
\begin{aligned}
    R_{1,2}=\dfrac{\cvp \varPhi^{2}(\alpha^2-1)}{2\gamma_{\infty} \alpha^{2/3}\left[\ln\left(\frac{1+\sqrt{1-1/\alpha^{2}}}{1-\sqrt{1-1/\alpha^{2}}}\right)\right]^{2}} 
    \left\{1\pm\sqrt{1+\dfrac{8\gamma_{\infty}\delta \alpha^2}{\cvp \varPhi^{2}(\alpha^2-1)}\left[\ln\left(\frac{1+\sqrt{1-1/\alpha^{2}}}{1-\sqrt{1-1/\alpha^{2}}}\right)\right]^{2}}\right\}.
    \label{eq:est_ineq_solR}
\end{aligned}
\end{equation}
\end{widetext}
For $R$ to be real, the argument of the large square root must be non-negative. The situation of negative argument can happen only when $\delta<0$.

\begin{figure}[!htb]
    \centering
    \subfloat[]{\includegraphics[width=0.8\linewidth]{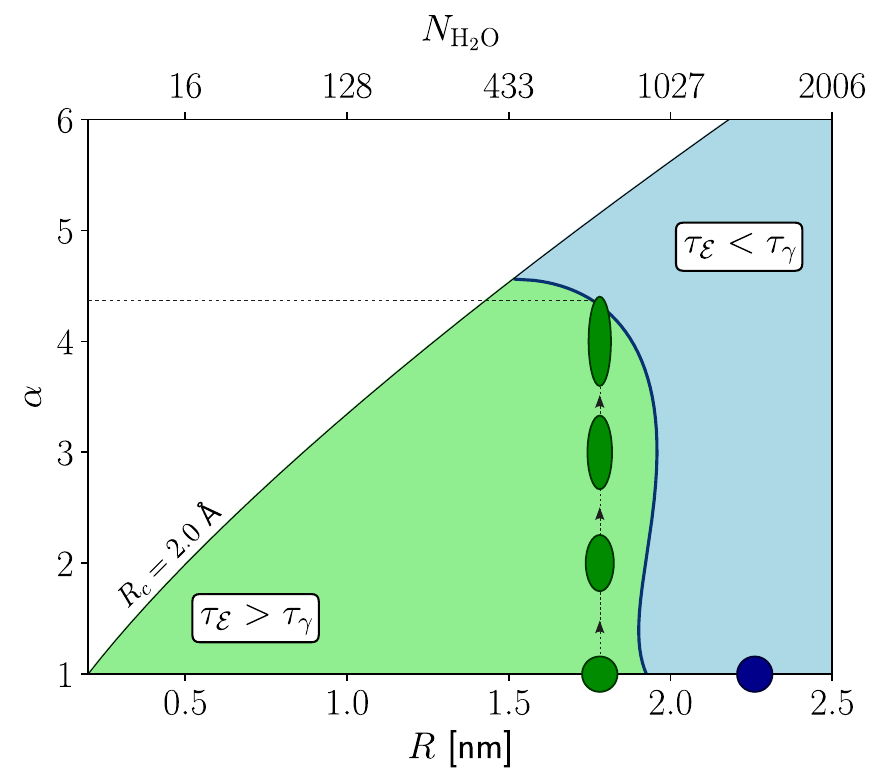}}
    
    \subfloat[\label{particle_est_graph_time}]{\includegraphics[width=0.8\linewidth]{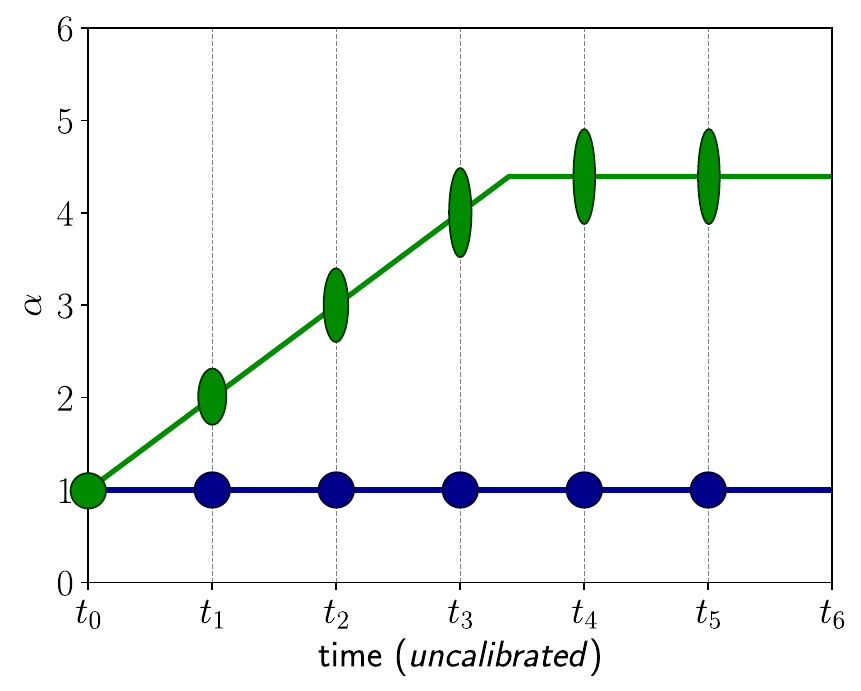}}
    \caption{(a) Plot of \cref{eq:est_ineq_solR} for $\varPhi=10$ V, $\gamma_{\infty}=\gamma_{SV}$, and $\delta=-0.5$ {\AA} and (b) time evolution of grains that are (green) or are not (blue) affected by electrostatic stresses. $R$ is assumed to be constant for the grains in the schematic.  $R_c=2$ {\AA} is set as a curvature size limit for water clusters.}
    \label{fig:particle_est}
\end{figure}

An example case for \cref{eq:est_ineq_solR} is plotted in \cref{fig:particle_est} using nominal parameters $\Phi =10$ V, $\gamma_{\infty}=\gamma_{SV}=109$ mN/m and $\delta=-0.5$ {\AA}. Solutions of \cref{eq:est_ineq_solR} are plotted up to $R_c=2$ {\AA}, which has been selected as representing the minimum possible radius of curvature for water clusters. This $R_c$ value is in agreement with the $R_c$ values observed in the \gls{md} simulations to be presented in \cref{sec:md_valid} (see \cref{fig:md_results_220K,fig:md_results_273K}). \textcolor{black}{It is also in agreement with} the smallest water cluster sizes in the \gls{qm} computations to be presented in \cref{sec:surf_tens} (see \cref{fig:surface_tension}). \textcolor{black}{An important constraint incorporated in the calculations is that} a grain cannot have an $R_c$ value smaller than the \textcolor{black}{nominal} molecular radius of a single water molecule, \textcolor{black}{which from experimental measurements and the reaction rate theory is $\sim 1.4$ {\AA}} \cite{Schatzberg1967}. The white region in \cref{fig:particle_est} corresponds to $R_c<2$ {\AA}, i.e., where $R$ solutions become nonphysical. Also, in \cref{fig:particle_est}, the solution of \cref{eq:est_ineq_solR} for which the positive sign has been chosen for the square root is plotted. The solution resulting from using the negative sign of the square root is located entirely in the white region, and since this region is nonphysical, it is not plotted. Results are also given with respect to the number of molecules in the grain (upper horizontal axis). The relation between the number of water molecules $N_{\ch{H2O}}$ and the radius $R$ is
\begin{equation}
N_{\ch{H2O}} = \dfrac{4\pi R^3 \rho_{ice}}{3 M_{H_2O}}
\label{eq:N_H2O}
\end{equation}
where $M_{H_2O}=18.015$ amu is the mass of a water molecule and $\rho_{ice}=917$ kg/m$^3$ is the density of ice \cite{Hobbs2010}.

Two regions are distinguished in \cref{fig:particle_est}, located on the left and right side of the $R$ solution of \cref{eq:est_ineq_solR}. Any particle with size and aspect ratio inside the green region satisfies \cref{eq:est_ineq} and thus its shape will be affected by electrostatic stresses. Therefore, a spherical grain ($\alpha\rightarrow1$) in the green region is unstable and will deform into an ellipsoid. On the other hand, electrostatic stresses cannot affect grains inside the blue area. For example, consider a spherical grain with $R=2.25$ nm. This case corresponds to the blue grain in \cref{fig:particle_est} located on the horizontal axis. Due to its relatively large $R$, this grain is within the blue region. Therefore, the shape of this grain will not change with time, i.e., it will remain spherical.

\begin{figure}[!b]
    \centering
    \includegraphics[width=0.8\linewidth]{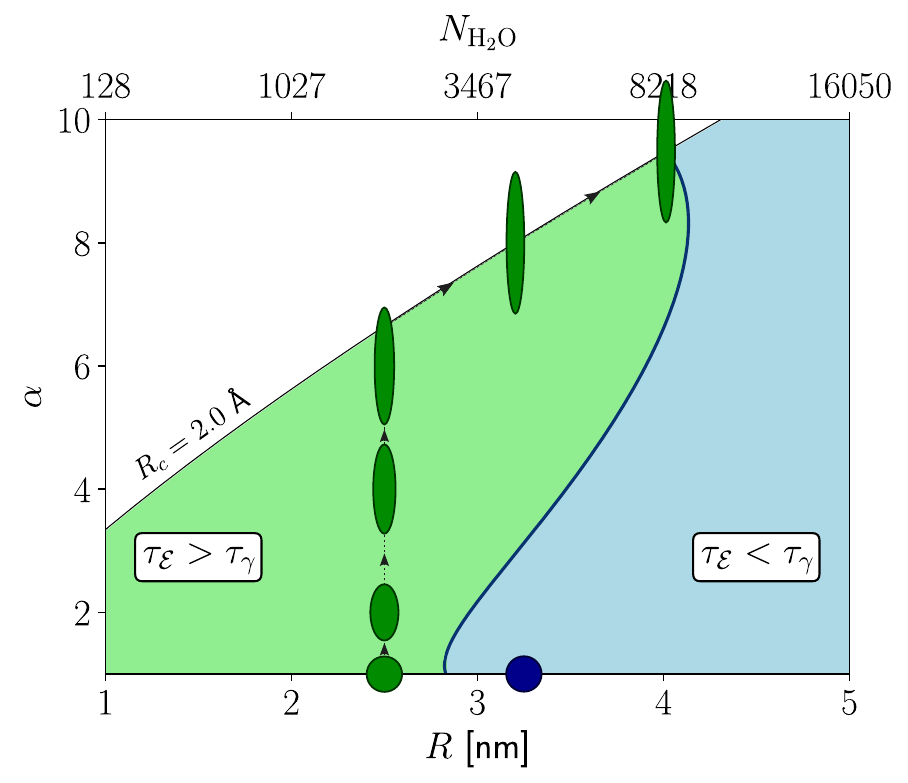}
    
    \caption{Plot of \cref{eq:est_ineq_solR} for $\varPhi=10$ V, $\gamma_{\infty}=\gamma_{LV}$, and $\delta=-0.4$ {\AA}.  $R_c=2$ {\AA} is set as a curvature size limit for water clusters.}
    \label{fig:particle_est_growth}
\end{figure}

\begin{figure*}[!ht]
    \centering
    \subfloat[$\gamma_{\infty}=0.109$ N/m]{\includegraphics[height=7.2cm]{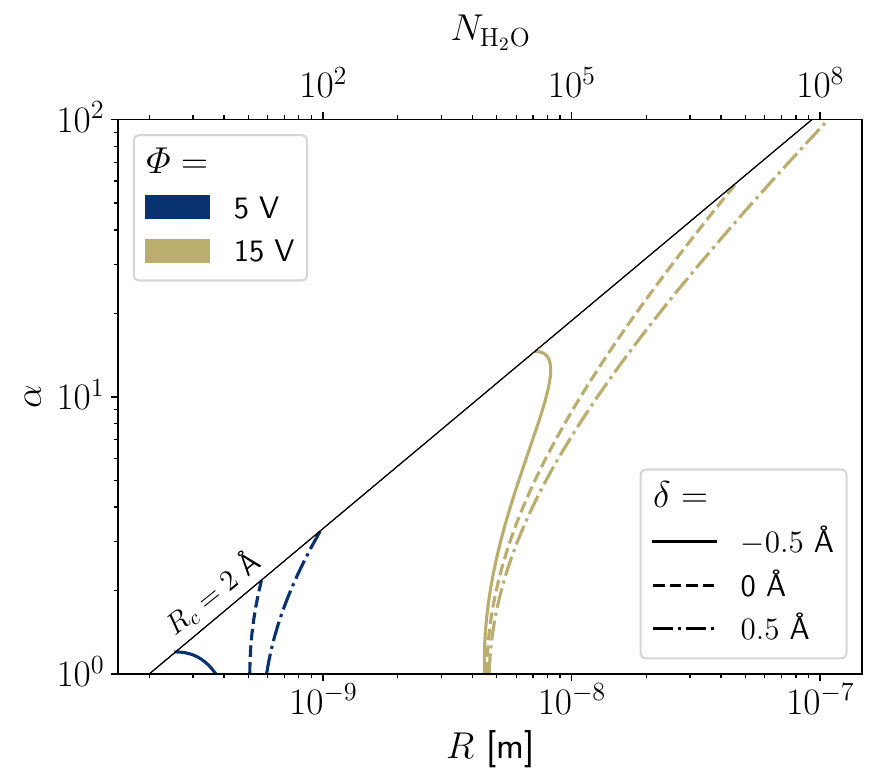}}\hspace{10pt}%
    \subfloat[$\delta=0$ {\AA}\label{fig:est_sol_large_delta0}]{\includegraphics[height=7.2cm]{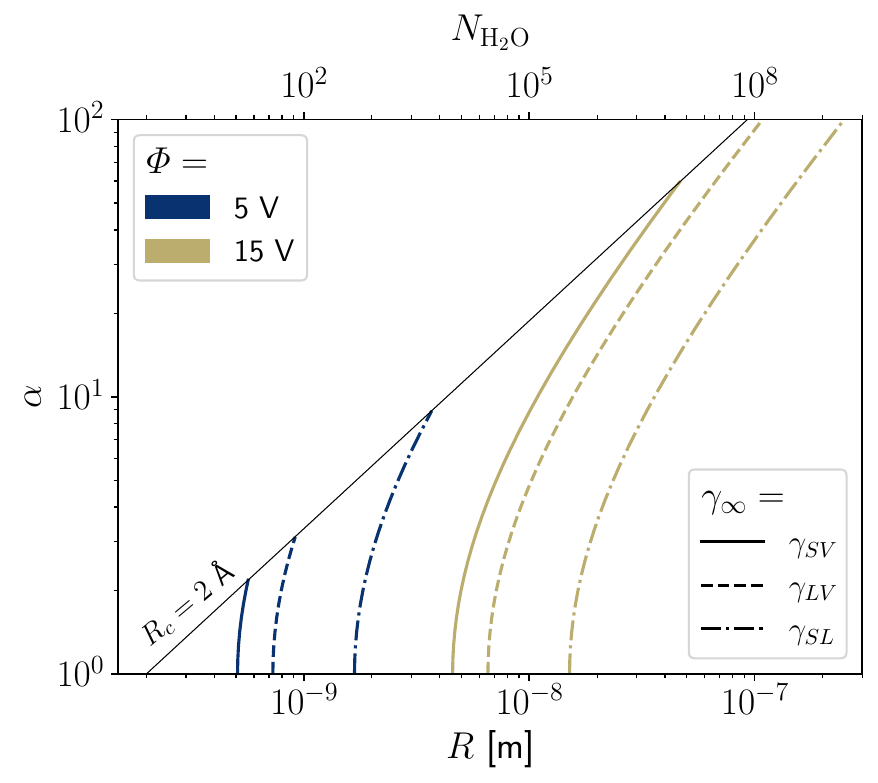}}
    \caption{Plot of \cref{eq:est_ineq_solR} for different material and plasma properties. The area to the left of each $R$ solution corresponds to $\tau_{\mathcal{E}} > \tau_{\gamma}$, whereas the area to the right corresponds to $\tau_{\mathcal{E}} < \tau_{\gamma}$. A curvature size limit of $R_c=2$ {\AA} is set for water clusters.  Thus, solutions with an $R_c$ value lower than the 2 {\AA} limit are not plotted. More lines are plotted in \cref{fig:est_sol_large_manylines}.}
    \label{fig:est_sol_large}
\end{figure*}

Now, consider an almost spherical grain with $R=1.75$ nm. This case corresponds to the green grain schematically shown in \cref{fig:particle_est}. Since its $R$ is within the green region, electrostatic stress will exceed the surface tension stress and deform the grain to become ellipsoidal. As long as the grain remains in the green area, it will continue to elongate. The time evolution of the aspect ratio of the grain is qualitatively depicted in \cref{particle_est_graph_time}. Assuming that during this deformation the grain does not interact with the plasma (i.e., $t_6$ is smaller than the grain--\ch{H2O} mean collision time), $R$ remains constant. Therefore, the \textcolor{black}{grain} aspect ratio will increase with time (i.e., move vertically in the $\alpha$--$R$ plane). At some point, $\alpha$ will reach its maximum value, since above this point, the surface tension of the grain increases considerably due to the Tolman correction \textcolor{black}{so} \cref{eq:est_ineq} is no longer satisfied. After the grain reaches its maximum aspect ratio, it will stop elongating.

Under certain conditions, electrostatic stresses may dominate for an ellipsoidal grain but not for a \textcolor{black}{same-volume} spherical grain. This is depicted in \cref{fig:particle_est_growth} where $\gamma_{\infty}=\gamma_{LV}=76$ mN/m is used. An initially spherical grain with $R=2.5$ nm (green grain) will become ellipsoidal under the influence of electrostatic forces. The elongation will \textcolor{black}{eventually} stop \textcolor{black}{because the radius of curvature at the tip becomes} very small. Now, suppose that the grain accretes some water molecules while maintaining its aspect ratio. This accretion of water molecules means that $R$ will increase and so the position in \cref{fig:particle_est_growth} will move slightly to the right (slightly larger $R$ at constant $\alpha$). The grain will then be slightly below the $R_c =2 $ {\AA} slanted line and so will again be in a situation where it is unstable to become more ellipsoidal. It will then move vertically in the figure until it reaches the $R_c =2 $ {\AA} line, but now at a larger $R$ and larger $\alpha$. With additional accretion of molecules, this process will continue, and the grain will creep up the $R_c =2 $ {\AA} slanted line. Eventually, the elongation will stop again due to the Tolman correction (reaching the blue region). At the end of this process, the green grain will have an $R$ value of $4$ nm and a 9.5 aspect ratio. However, an initially spherical grain with $R=3.25$ nm (blue grain) will never access the green region in \cref{fig:particle_est_growth} and thus will remain a sphere.

\Cref{fig:est_sol_large} shows solutions of \cref{eq:est_ineq_solR} for different values of $\Phi$, $\delta$, and $\gamma_{\infty}$. A broader range of properties is plotted in \cref{fig:est_sol_large_manylines}. Overall, two size-dependent main regions can be distinguished in which electrostatic stress may or may not overcome the surface tension stress. If the initial radius of a nearly spherical grain is sufficiently large (e.g., initial radius greater than 1 \micro m), the inequality of \cref{eq:est_ineq} will never be satisfied. The reason is that a large spherical grain has too few electrons per unit surface area compared to a small grain charged to the same potential. This is seen by considering that $\sigma= Q/(4\pi R^2)$ and $Q=4\pi\cvp R\varPhi$ so $\sigma=\cvp \varPhi /R$. Thus, electrostatic stress scales as $\Phi^2/R^{2}$, whereas, ignoring the effects of Tolman's correction, surface tension scales as $1/R$.   Therefore, electrostatic stresses are negligible on spherical grains having a large initial radius.

Nanosized grains can be affected by internal electrostatic stresses, depending only on the surface tension of the material and the \textcolor{black}{floating} potential. The surface tension of nanosized curvatures (i.e., with $R_c\approx 1$ nm) depends strongly on Tolman's correction. Therefore, whether the inequality of \cref{eq:est_ineq} will be satisfied depends on the sign and magnitude of Tolman's length. A positive $\delta$ aids electrostatic stresses (cf. \cref{fig:est_sol_large} and \cref{fig:est}), and thus the inequality can be satisfied at larger grain sizes or lower \textcolor{black}{floating} potentials. The inequality may never be satisfied if $\delta$ has a negative value. The reader is referred to the findings in \cref{sec:surf_tens} showing a negative $\delta$ value for water clusters. Once the inequality of \cref{eq:est_ineq} is satisfied, an initially spherical grain will evolve into an ellipsoid. As aforementioned, the elongation will continue until atomistic effects on the tip of the ellipsoid become significant (i.e, small $R_c$ or increased surface tension due to a negative Tolman length).

After nanosized grains become ellipsoidal, they can continue growing into more elongated shapes due to external factors even if electrostatic stresses become negligible. The two tips of the ellipsoid are sharply pointed (i.e., have a small radius of curvature) and so have strong local electric field gradients. Water molecules have a large dipole moment $\pmb{\mu}$ and thus incoming water molecules will be attracted to the tips of the ellipsoid due to the polarization force $\bm{F}=(\pmb{\mu}\cdot\nabla)\pmb{\mathcal{E}}$ \cite{Griffiths2013}. The dipole moment $\mu$ of a gaseous water molecule is 1.85 D \cite{Lovas1978}. As a consequence, more water molecules will accumulate at the two ends of the ellipsoid, resulting in a more elongated shape \cite{Bellan2020}. This can explain why experimental studies observe elongated macrosized grains with a large aspect ratio\cite{Chai2018,Nicolov2024}.

Additional parameter scans showing these behaviors are given in the Supporting Information. In particular, the two sides of  \cref{eq:est_ineq} are plotted on \cref{fig:est} for a range of parameters. The solution of \cref{eq:est_ineq} with respect to the aspect ratio, $\alpha$, is also given in \cref{sec:si_elec_surf_stress}.

\section{Validation using molecular dynamics}
\label{sec:md_valid}

\begin{figure*}[!htb]
    \centering
    \includegraphics[width=0.6\linewidth]{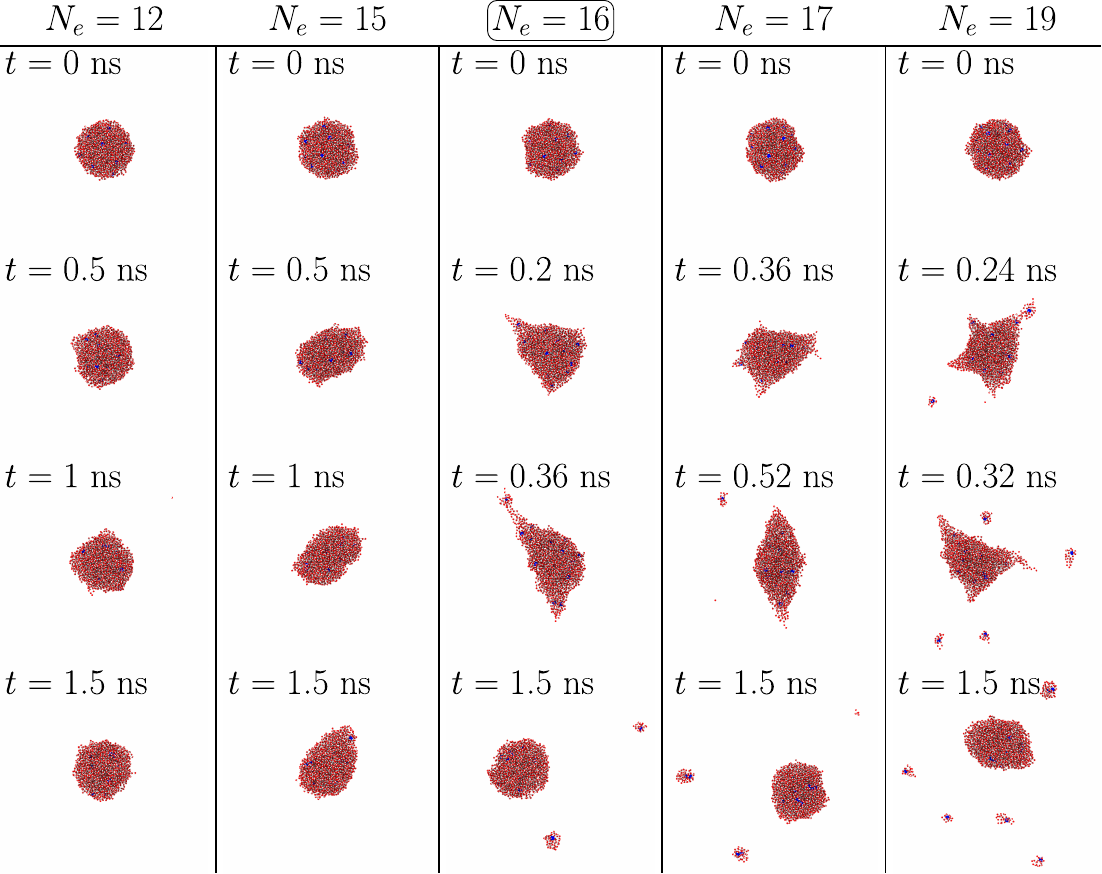}
    \caption{Atomic trajectories of grains with different numbers of electrons at $T=273$ K. Oxygen atoms are in red; hydrogen atoms are in gray; electrons are in blue. High-resolution images can be found in the online version.}
    \label{fig:md_results_273K}
\end{figure*}

The theoretical analysis of \cref{sec:elec_surf_stress} is now validated using \gls{md} simulations. \gls{md} is selected since it can simulate the dynamics and kinetics of many thousands of atoms.  \gls{md} simulations are more suitable than \gls{qm} computations here since \gls{qm} computations are typically limited to less than a few hundred atoms (cf. \cref{sec:solv_cond}). It is noted that the electronic structure of atoms is not described in \gls{md} simulations, and all atoms and electrons are simulated as particles. \gls{qm} computations would offer a higher accuracy, but the computational cost of simulating the whole grain is unfeasible. To achieve good accuracy with \gls{md}, the bonded and non-bonded interactions among the particles are calibrated using \gls{qm} computations. Therefore, a high degree of accuracy can be achieved while retaining a significantly lower computational cost compared to \gls{qm} simulations. The calibration of the \gls{md} simulations of this study is detailed below.

Spherical grains with a radius of 25~{\AA} and different numbers of electrons are investigated. Using \cref{eq:N_H2O} and the density of water, $\rho_{water}=997$ kg/m$^3$, this radius corresponds to 2181 water molecules.  The grains are constructed with the charge distribution of spherical conductors. Therefore, electrons are uniformly placed close to the surface of the grain. Specifically, electrons are placed on a 20 {\AA} radius spherical surface that is concentric with the spherical grain. The Fibonacci sphere sequence was used to place the electrons uniformly, which is detailed in \cref{sec:si_valid_md}.

The lowest potential for which the inequality of \cref{eq:est_ineq} is satisfied for a spherical grain ($\alpha \rightarrow 1$) is
\begin{equation}
    \varPhi_{sph,lim} = \sqrt{\dfrac{4 \gamma_{\infty}R^2}{\cvp(R+2\delta)}}.
\end{equation}
From \cref{eq:q}, the corresponding charge of a spherical grain with $\varPhi=\varPhi_{sph,lim}$ is
\begin{equation}
    Q_{sph,lim}= 8\pi R^2
    \sqrt{\dfrac{\cvp\gamma_{\infty}}{R+2\delta}}.
    \label{eq:q_sph_lim}
\end{equation}
\textcolor{black}{If we neglect Tolman's correction (i.e., set $\delta=0$), then \cref{eq:q_sph_lim} simplifies to Rayleigh’s charge limit}\cite{Rayleigh1882,Kebarle2000}.

For the parameters of \cref{tab:q_d_param} and assuming a $\delta$ value of $-0.5$ {\AA}, the lowest potential for which \cref{eq:est_ineq} is satisfied is 9.46 V for liquid (using $\gamma_{\infty}= 0.076$ N/m) and 11.32 V for solid \ch{H2O} grains (using $\gamma_{\infty}= 0.109$ N/m). The corresponding charge is 16.42 and 19.66 electrons for liquid and solid grains, respectively. Values for $\delta=0$ {\AA} are given in \cref{tab:phi_q_limit}. The $\alpha-R$ curves for liquid grains and $\delta$ values of $-0.5$ and 0 {\AA} are shown in \cref{fig:est_sol_large_delta_md_case}.

\begin{figure*}[!tb]
    \centering
    \includegraphics[width=0.6\linewidth]{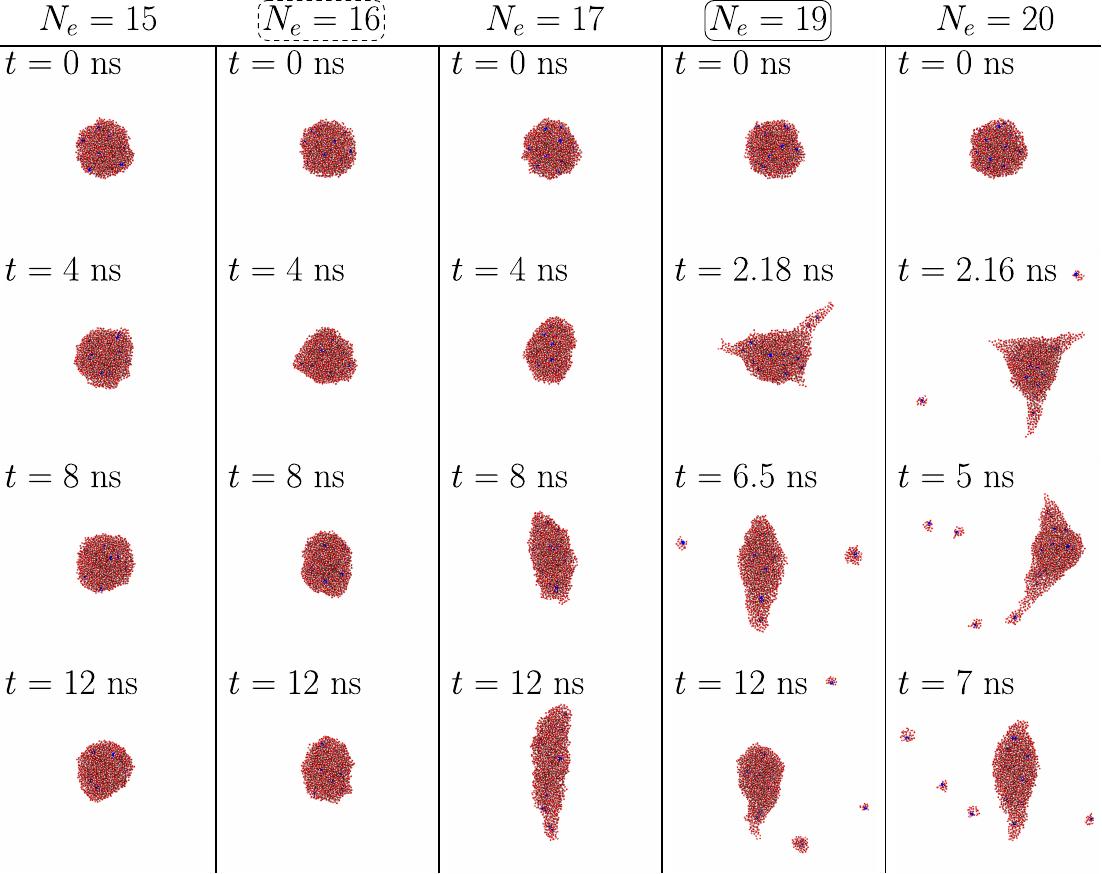}
    \caption{Atomic trajectories of grains with different numbers of electrons at $T=220$ K. Oxygen atoms are in red; hydrogen atoms are in gray; electrons are in blue. High-resolution images can be found in the online version.}
    \label{fig:md_results_220K}
\end{figure*}

It is noted here that these simulations are performed under a constant charge, whereas in the analysis of \cref{sec:elec_surf_stress}, the grains have a constant electric potential. Simulations of grains at constant charge are valid if the grain deformation is faster than the rate at which electrons or ions collide with the grain and so change the charge on the grain.  The lower limit of the grain charging time, $\tau_{q}$, is found by considering only electrons impacting a spherical grain of radius $r_d$. The electron current density to the grain is $J_e= e n_e v_{T_e}$, where the electron thermal velocity $v_{T_e}=\sqrt{2\kappa T_e/m_e}$. The electron current to a spherical grain is $I=4\pi r_d^2 J =Q/\tau_q$. Therefore, using \cref{eq:q}, the charging time is
\begin{equation}
    \tau_q=\dfrac{\cvp \varPhi}{e r_d n_e v_{T_e}}.
    \label{eq:tau_q_2}
\end{equation}
The above equation is a lower limit for the charging time since repulsions of electrons by the charge of the grain and collection of ions will slow down the charging process. The charging time scales as $r_d^{-1}$, and thus smaller particles take longer to charge. Assuming that $\varPhi\approx2\kappa T_e/e$ (c.f., \cref{fig:Phi_parametric}) the charging time becomes
\begin{equation}
    \tau_q=\dfrac{v_{T_e}}{r_d \omega_{pe}^2},
    \label{eq:tau_q}
\end{equation}
where $\omega_{pe}$ is the electron plasma frequency given by $\omega_{pe}^2=n_e e^2/(\cvp m_e)$. For the grain considered in this section with $r_d=25$ {\AA}, the conditions of \cref{tab:q_d_param}, \textcolor{black}{and assuming $n_e=n_i$}, $\tau_q\approx 100$ \micro s. This is a short time compared to astrophysical situations and the Caltech experiment, but a long time compared to the few nanoseconds characteristic time of the \gls{md} simulations. The assumption that the charge is constant in the \gls{md} simulations is thus appropriate.

\begin{table}[!htb]
    \centering
    \begin{tabular}{c|cccc}
        Element & m [amu] & q [e] \\
        \hline
       \ch{O}   & 15.9994 & $-0.8476$ \\
       \ch{H}   & 1.00794 & $0.4238$ \\
       \ch{$e$}  & 1.00794 & $-1.0$ \\
       \hline
    \end{tabular}\hspace{6mm}%
    \begin{tabular}{c|cc}
        Pair & $\sigma_{LJ}$ [{\AA}] & $\epsilon_{LJ}$ [kcal/mol] \\
        \hline
       \ch{O-O}   & 0.15535 & 3.166 \\
       \ch{$e$-H}  & 1.46998 & 2.052 \\
       \ch{$e$-O}  & 2.36088 & 0.232 \\
       \hline
    \end{tabular}
    \caption{\gls{md} simulation details. The H and O values are based on the SPC/E model. The model for \ch{$e$-H} and \ch{$e$-O} interactions is detailed in \cref{sec:si_valid_md}. The \ch{OH} bond distance is set to 1.0 {\AA} and the $\angle\ch{HOH}$ is 109.47$^{\circ}$. The LJ values for the \ch{H-O}, \ch{H-H}, and $e$--$e$ interactions are zero.}
    \label{tab:lj_values}
\end{table}

In this study, the SCP/E water potential \cite{Berendsen1987} is used since it can accurately predict the surface tension of liquid water compared to experimental measurements \cite{Chen2007}. The \gls{lj} potential parameters for solvated electrons are developed according to the methodology discussed in \cref{sec:si_valid_md}. All simulation parameters are given in \cref{tab:lj_values}. Also, the potentials for the \ch{$e$-H} and \ch{$e$-O} interactions are plotted in \cref{fig:pot_lj}.  The \gls{md} model predicts with sufficient accuracy the cavity size of solvated electrons compared to \gls{qm} computations  (cf. \cref{sec:si_valid_md}). A more accurate description of solvated electrons in \gls{md} simulations is the scope of future work. It is noted that the electron mass is assumed to be equal to the mass of the hydrogen atom. This assumption, which is followed by other \gls{md} studies\cite{Su2009,Islam2016}, allows for a significantly longer timestep and thus a lower computational cost.

To investigate the stresses on both liquid and solid grains, \gls{md} simulations are performed at 220 K and 273 K. At temperatures above this range, rapid evaporation and dissociation of the grain may occur. On the other hand, at lower temperatures, the electrons and molecules move too slowly to observe any changes within a few nanoseconds. Simulation details are given in \cref{sec:si_valid_md}. Simulations at 273 K were terminated at 1.5 ns because the electrostatic stress effects were evident within this timeframe. The total simulated time for the simulations at 220 K was 12 ns due to the high computational cost. For instance, a 2ns simulated time at 220 K took approximately 14.4 hours to run on 16 processors. The time evolution of the grain structures is shown in \cref{fig:md_results_273K,fig:md_results_220K} for different numbers of electrons. It is noted that the simulations performed at 220 K may not be precise. The reason is that, even though the SCP/E model can accurately predict the surface tension of liquid water, it has not been tested for ice to the best of the authors' knowledge.

First, consider the simulations of liquid grains at 273 K shown in \cref{fig:md_results_273K}. Grains with only 12 electrons retain their spherical shape since the surface tension completely dominates any electrostatic forces. As the grain charge approaches 16 electrons,  electrostatic and surface tension stresses become of the same order of magnitude, and the grain becomes slightly ellipsoidal. When the equality of \cref{eq:est_ineq} is satisfied (i.e., $N_e=16$), the shape of the grain is significantly affected by the mutual repulsions of the electrons with each other. As a result, the structure of the grain changes drastically. This change happens too fast, leading to the disintegration of the 273 K grain into separate smaller parts instead of a more steady elongation. The grain has lost an electron at the end of the 1.5 ns. \Cref{eq:est_ineq} is no longer satisfied; thus, the grain returns to a more spherical shape. As the number of electrons increases, deformation and dissociation of the grain occur earlier and with more free hydrate clusters present. Based on the results of \cref{fig:md_results_273K}, it is concluded that the \gls{md} simulations and the theory of \cref{sec:elec_surf_stress} are in agreement within a single-electron accuracy.

The influence of electrostatic forces is also observed in the simulations of ice grains at 220 K shown in \cref{fig:md_results_220K}. Due to the slower movement of water molecules and electrons, a longer simulated time is needed to observe changes in the structure of the grains compared to the simulations at 273 K. Ice grains become ellipsoids under the influence of electrostatic stresses. For example, in the case of $N_e=17$ at $t=12$ ns, fitting an ellipse to the grain gives $\alpha$ of approximately 3.4 to 3.5. Interestingly, the 220 K  grains retain their ellipsoidal shape compared to the 273 K grains, even if smaller hydrate clusters dissociate and escape. The final electron distribution of the ellipsoidal grains is fairly uniform along the z-axis, which is in accordance with \cref{eq:z_d_z_ell}. These \gls{md} results for ice agree with the analysis in \cref{sec:elec_surf_stress} within good accuracy. There is a deviation of a couple of electrons, which likely occurs due to the fact that the $\gamma_{SV}$ value predicted by SPC/E may differ from the experimental measurements.

\gls{md} simulations showed that \textcolor{black}{spherical solid} grains break apart, \textcolor{black}{when their charge exceeds the $Q_{sph,lim}$ limit given in \cref{eq:q_sph_lim}, even} before they reach their maximum possible aspect ratio (cf. \cref{fig:md_results_273K,fig:md_results_220K,fig:est_sol_large}). However, ellipsoidal growth \textcolor{black}{of these grains} may still happen in \textcolor{black}{astrophysical dusty plasmas}. \textcolor{black}{The lifetime of astrophysical grains is seconds to millions of years, but the ion-grain and electron-grain collision times are many orders of magnitude smaller. In those cases,} the grains \textcolor{black}{do not remain isolated and can be thought of as being} continuously bombarded by electrons. Therefore, even if \textcolor{black}{astrophysical} grains \textcolor{black}{break apart and} lose electrons, they will acquire new ones. As a consequence, it is expected that their aspect ratio will increase.

\section{Stresses on uniformly charged grains}

\cref{sec:solv_cond} will show that at small timescales, the water-based grains cannot be considered conducting and thus would not necessarily have the charge distribution discussed in \cref{sec:elec_surf_stress,sec:md_valid}. The reason is that if the grain is an insulator, the charge distribution can be arbitrary and is not limited to $Q=C\varPhi$. In that case, \cref{eq:elec_stress} should be solved for the appropriate surface charge density and thus \cref{eq:stresses_ineq} can only be simplified to
\begin{equation}
    \dfrac{R+2\alpha^{4/3}\delta}{2\alpha^{4/3}} \ge \dfrac{\cvp\gamma_{\infty}}{\sigma^2}.
    \label{eq:est_unif}
\end{equation}
To investigate a different charge distribution, below, we focus on the case of a grain with a uniform charge distribution.

The work to create a uniform spherical charge distribution is \cite{Griffiths2013}
\begin{equation}
    W_{unif,sph} = \dfrac{3 Q^2}{20 \pi \cvp R}.
    \label{eq:work_unif}
\end{equation}
Comparing \cref{eq:work_unif,eq:work_cond}, it can be seen that the work of a sphere with a uniform charge distribution is only 1.2 times greater than that of a conducting sphere. Therefore, a uniform distribution is not an impossible configuration. Eventually, a uniform distribution will relax to a conductor distribution since the latter is in a lower energy state.

At the tip of the ellipsoid, the surface charge density of a uniform charge distribution is
\begin{equation}
\begin{aligned}
    \sigma_{unif}(a) = \dfrac{3 Q}{4 \pi \alpha^{4/3}R^2}.
    \label{eq:sigma_unif}
\end{aligned}
\end{equation}
The ratio of $\sigma_{unif}(a)$ to $\sigma_{cond}(a)$ (i.e., \cref{eq:sigma_unif,eq:sigma_cond}) is $3/\alpha^{2}$. Hence, a uniformly charged grain with $\alpha \ge \sqrt{3}$ needs to be charged with significantly more electrons compared to a conducting grain in order to satisfy \cref{eq:est_unif} and be deformed.

\section{Solvated electrons and conductivity}
\label{sec:solv_cond}

In this section, the binding nature of solvated electrons in water clusters is investigated and compared to electron-electron repulsions. Solvated electrons are bound to a solvent similarly to a solvated anion. Electrons dissolved in water are also known as hydrated electrons. Solvated electrons must be present within the ice grain structures since the grains are continuously bombarded by electrons that may not have enough energy to ionize the water molecules \cite{Cottin1959,Fedor2006}. Additionally, solvated electrons have been experimentally identified using Electron Paramagnetic Resonance (EPR) measurements \cite{Kevan1981} in water at $T_g=77$ K, which is a temperature similar to that of astrophysical dusty plasmas \cite{Shukla2001,Woitke2009,Nicolov2024}. 

\begin{figure}[!htb]
    \centering
    \includegraphics[width=0.85\linewidth]{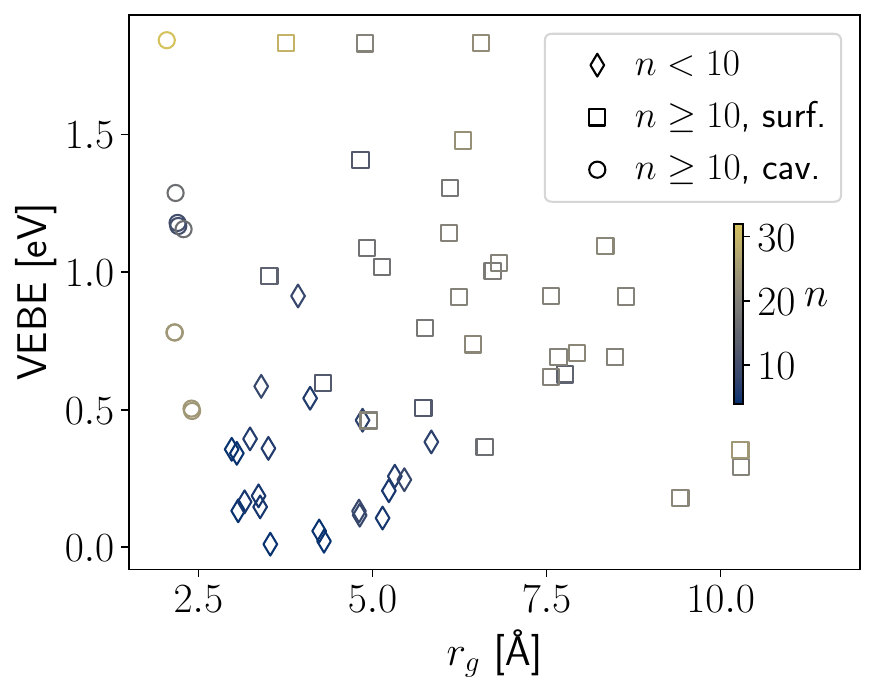}
    \includegraphics[width=0.9\linewidth]{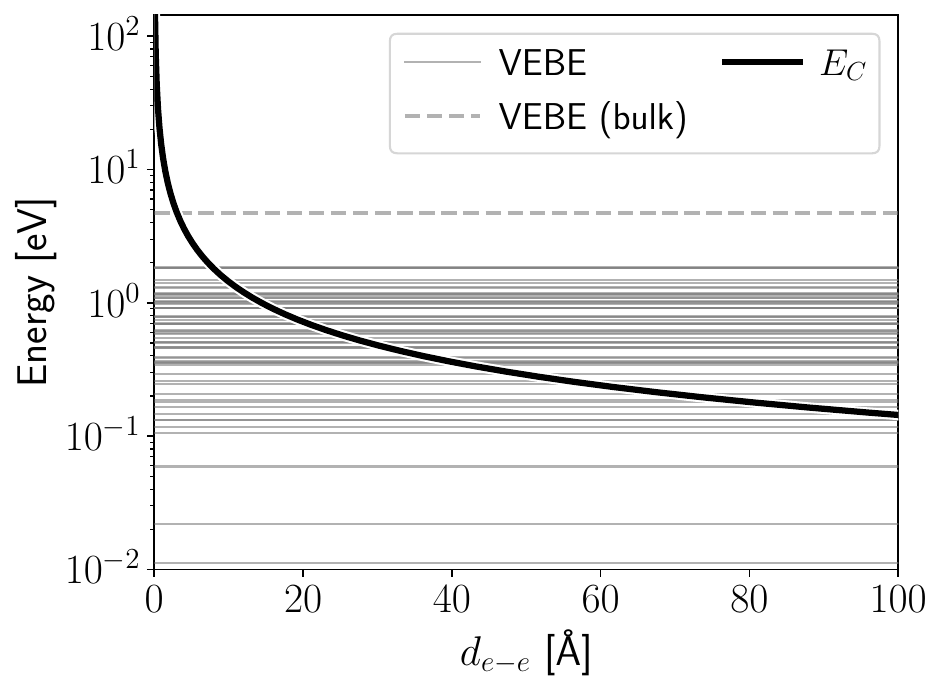}
    \caption{VEBE of clusters with different numbers of water molecules with respect to their radius of gyration (top) and comparison of the VEBE values with the Coulomb repulsions between two electrons (bottom). In the top figure, different colors and symbols distinguish the cluster size and the type of solvated electrons (i.e., surface or cavity). The VEBE value of the bulk material is taken from Ref. ~\citenum{Thurmer2021}.}
    \label{fig:cond_vebe}
\end{figure}

When an electron is injected into an aqueous cluster, the electron does not get solvated immediately but rather exists in the conduction band (i.e., the lowest unoccupied band of energy levels) of the solvent \cite{Palianov2014,Pizzochero2019}. During this initial stage, the electron behaves as a quasi-free electron and can have a delocalization length of approximately 40 {\AA}.\cite{Savolainen2014} When a broken water bond is found, the electron gets solvated and forms a cavity. Specifically, the electron gets stabilized and trapped by an electrostatic potential created by the dipole moments and the dielectric response of the surrounding water molecules \cite{Jortner1964,Kumar2015}. In water, the trapping of the solvated electron takes less than half a picosecond \cite{Pizzochero2019}.

Solvated electrons can be categorized into two groups: the surface electrons, which are localized on the surface of the clusters, and the cavity electrons, which are encapsulated by water molecules. Surface electrons are typically bound with three or fewer water molecules \cite{Herbert2006}. Cavity electrons are located within a water structure and are bound with four \cite{Kumar2015}, five \cite{Ambrosio2017}, or six \cite{Kevan1981} nearby water molecules. Cavity electrons interact with the dipole moment of the OH bond of the water molecules \cite{Kumar2015}, but surface electrons can also interact with the total \ch{H2O} dipole moment \cite{Zhan2003}.

A measurement of the volume that the solvated electron occupies is the radius of gyration $r_g$ which is defined as
\begin{equation}
\begin{aligned}
    r_g = \sqrt{\dfrac{\int r^2 \csd{\bm{r}} \dd \bm{r}}{\int \csd{\bm{r}} \dd \bm{r}}},
    \label{eq:rg_qm}
\end{aligned}
\end{equation}
where $\rho_s = |\rho_\uparrow - \rho_\downarrow|$ is the spin density. $\rho_\uparrow$ and $\rho_\downarrow$ are the densities of electrons with opposite spins, i.e., spin-up and spin-down electrons, respectively. The radius of gyration of water clusters with solvated electrons is shown \cref{fig:cond_vebe}. Details about the geometries are given in \cref{sec:si_solv_cond}. Surface electrons have a larger $r_g$ value since they are only partially trapped by a couple of water molecules. Cavity electrons have $r_g \approx2$ {\AA}. Moreover, the \gls{csdf} is defined as:
\begin{equation}
    \mathrm{CSDF} = \dfrac{\int\bm{r}\csd{\bm{r}} \dd \bm{r}}{\int\csd{\bm{r}} \dd \bm{r}}.
    \label{eq:csdf}
\end{equation}

Next, we assess the assumption that water grains are electrically conducting. This is an important assumption that is presumed by the \gls{oml} theory. Specifically, we analyze how the electrons inside the grain would behave due to close electron-electron repulsions. The electron-electron Coulomb repulsion energy $E_C$ is
\begin{equation}
    E_C = \dfrac{e^2}{4\pi \cvp d_{e-e}},
    \label{eq:E_C}
\end{equation}
where $d_{e-e}$ is the distance between the two electrons.

A measurement of how strongly a solvated electron is bound to the water cluster is given by the \gls{vebe}, or work function. The \gls{vebe} is given by
\begin{equation}
    \mathrm{VEBE}=E_{\mathrm{neutral}}^{\mathrm{cavity}} - E_{\mathrm{ion}},
\end{equation}
where $E_{\mathrm{ion}}$ is the energy of the water cluster with a solvated electron and $E_{\mathrm{neutral}}^{\mathrm{cavity}}$ is the energy of the same geometry without the electron. In this study, \glspl{vebe} are computed using two \gls{qm} methods: the \gls{dft}\cite{Kohn1965} and the \gls{mp2} \cite{Moller1934}. These methods approximately solve the time-independent Schr{\"o}dinger equation. As a result, the electronic structure and potential energy of a molecular complex can be computed with accuracy. Due to their substantial computational cost, the \gls{qm} computations of this study are limited to water clusters with less than a hundred molecules.

\glspl{vebe} for different clusters are plotted in \cref{fig:cond_vebe}. Details on the geometries and simulations are given in \cref{sec:si_solv_cond}. Geometry optimizations were performed at the \gls{dft}/B3LYP-D3/6-311++G** level of theory, while the reported energies were found at the \gls{mp2}/6-311++G** level of theory. The reason is that, unlike \gls{dft}, the \gls{mp2} method can accurately capture the energy of solvated electron structures compared to \gls{ccsdt} \cite{Herbert2005,Herbert2006}. Results with negative VEBE values have been excluded from the analysis. Negative VEBEs may arise because the finite basis electronic wave function prohibits the solvated electron from escaping the water cluster \cite{Herbert2005}.

For most cavity structures, especially with $n \ge 10$, the \gls{vebe} gets lower than the Coulomb repulsions above distances of a few nanometers. These results suggest that ice grains cannot be considered electrical conductors for the nanoscale clusters investigated here. Note, however, that this scale is important since the accretion and growth processes take place at the nanosecond timescale. This proves that the \gls{oml} theory cannot be applied to small grains with a nanosecond existence. In those short scales, the actual charge distribution can be arbitrary. It would depend on several factors, such as the diffusion of electrons and water molecules within the grains and collisions with the plasma particles.

\begin{figure}[!htb]

    \subfloat[$\ch{(H2O)^{1-}_24}+\ch{(H2O)^{1-}_24}$ with a 12.83 {\AA} distance between the \glspl{csdf} of the two $e$.\label{fig:n24_24_spindens}]{\includegraphics[width=0.6\linewidth]{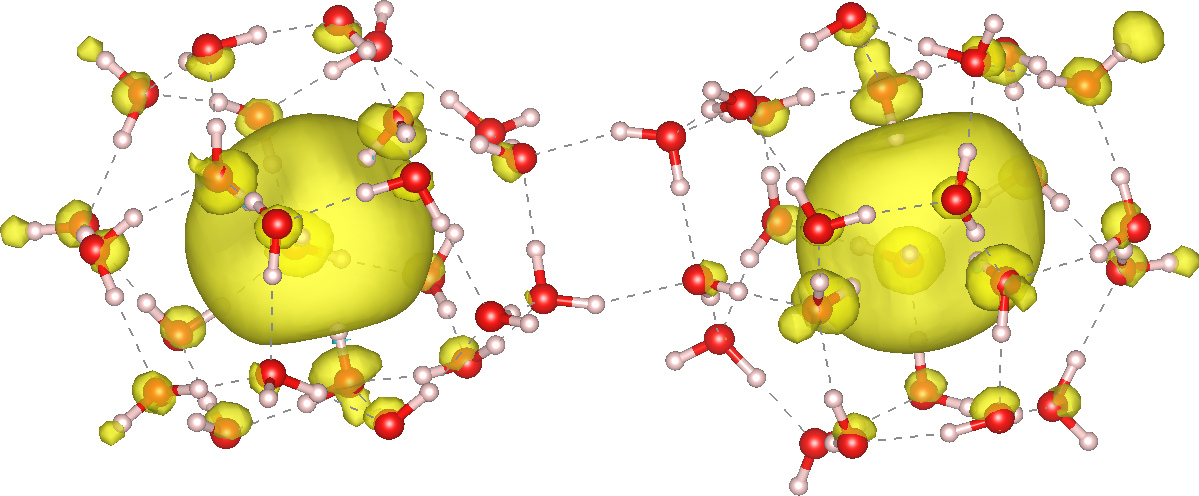}}
    \vspace{2mm}
    \subfloat[$\ch{(H2O)^{1-}_24}+\ch{(H2O)_21}+\ch{(H2O)^{1-}_24}$ with 18.13 {\AA} distance between the \glspl{csdf} of the two $e$.\label{fig:n24_21_24_spindens}]{\includegraphics[width=0.8\linewidth]{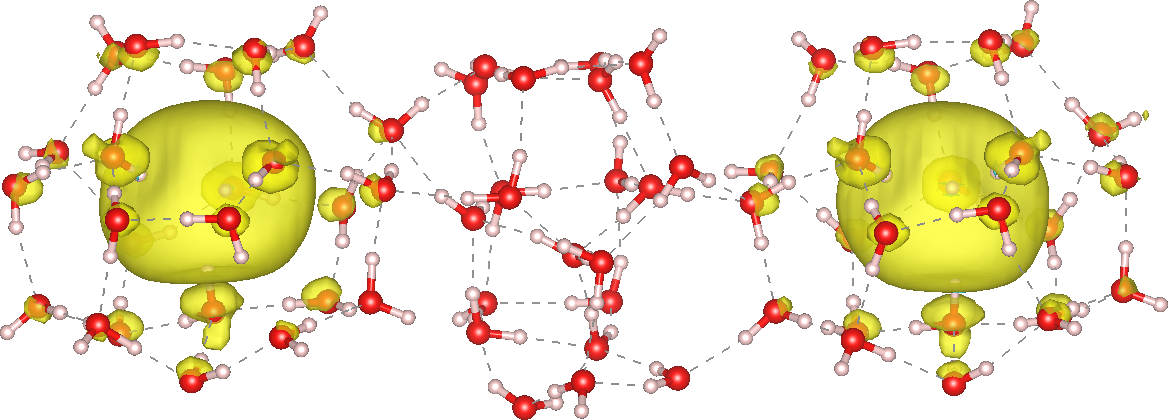}}

    \caption{AIMD simulations of different hydrate geometries with two solvated electrons. Results are presented at $t=500$ fs. Oxygen atoms are in red; hydrogen atoms are in gray; spin density is in yellow.}
    \label{fig:spindens}
\end{figure}

To validate the above analysis, we examine whether repulsions become negligible above a certain interelectron distance using \gls{aimd} simulations. Specifically, water clusters with two solvated electrons are simulated. The \gls{aimd} simulations are performed at the \gls{dft}/B3LYP-D3/6-311++G** level of theory. Due to the high computational cost, simulations are performed for a total of 500 fs simulation time. For instance, the total simulation time for the system shown in \cref{fig:n24_21_24_spindens} was approximately 14 days, running on 168 processors.

\Cref{fig:spindens} shows the spin density of two solvated electrons in two different configurations. Each solvated electron is within an $n=24$ water configuration with a \gls{vebe} equal to 0.78 eV. Given this energy value, \cref{eq:E_C} predicts a minimum electron-electron distance of 18.44 {\AA} for electrostatic interactions to be significant. In the two geometries of \cref{fig:spindens}, the \glspl{csdf} (given in \cref{eq:csdf}) of the two electrons are at distances of 12.83 {\AA} and 18.13 {\AA}. Therefore, electrons would only be significantly affected by electrostatics in the \cref{fig:n24_24_spindens} configuration. Indeed, the wavefunctions of the two electrons in \cref{fig:n24_24_spindens} are shifted outwards from the center-of-mass of the cluster. On the contrary, the wavefunctions in \cref{fig:n24_21_24_spindens} are not significantly affected by electrostatic repulsions and thus retain their initial structure.

Despite the above analysis proving that the conductivity assumption is invalid at the nanoscale, we now show that large macroscale grains can behave as electrical conductors. As aforementioned, \gls{qm} and \gls{md} methods cannot reach the necessary simulation time and number of molecules to investigate macroscale grains. Therefore, the conductivity assumption for macroscale grains is investigated theoretically.

On defining the electrical conductivity to be $\xi$ so the electric current density is $\bm{J}=\xi \pmb{\mathcal{E}}$, the characteristic time to achieve the charge distribution of a conducting material can be computed from Ampere's law, 
\begin{equation}
   \nabla \cross \bm{B} = \mu_0 \left(\xi \pmb{\mathcal{E}} +\cvp \dfrac{\partial \pmb{\mathcal{E}}}{\partial t} \right).
\end{equation} 
The magnetic field is assumed to be negligible. Therefore, the \gls{rhs} of Ampere's law gives the characteristic time for any initial electric field to decay to be $\tau_c = \cvp / \xi$.

\begin{figure}[!tb]
    \centering
    \subfloat[\label{fig:grain_conductivity_scheme_1}]{\includegraphics[width=0.66\linewidth]{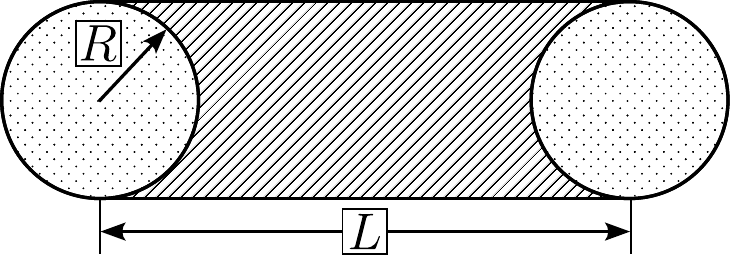}}

    \subfloat[\label{fig:grain_conductivity_scheme_2}]{\includegraphics[width=0.594\linewidth]{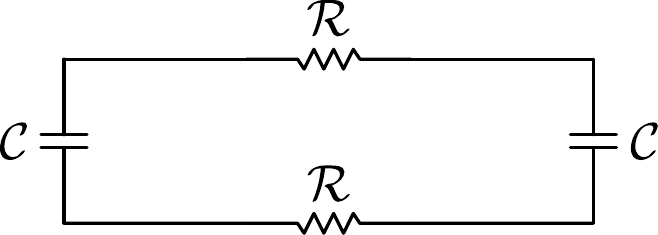}}
    
    \caption{(a) Theoretical model of an ice grain system, i.e., two spherical capacitors connected by a cylindrical resistor, and (b) equivalent circuit.}
    \label{fig:grain_conductivity_scheme}
\end{figure}

This characteristic time can equivalently be computed from the point of view of a tangible electric circuit by approximating the ellipsoidal grain as a resistor-capacitor (RC) circuit. Consider the system sketched in \cref{fig:grain_conductivity_scheme_1}. The ice grains are approximated as two spherical capacitors connected to each other by a cylindrical resistor. It is assumed that initially the left hand sphere is charged and the right hand sphere is uncharged so with time, charge will flow from the left hand sphere to the right hand sphere until half the charge is on each sphere. The effective circuit for this flow of charge is shown in \cref{fig:grain_conductivity_scheme_2}. The characteristic time of this circuit provides an estimate time for the charges in the grain to reach the expected distribution for a conducting material.

\begin{figure*}[!htb]
    \centering
    \subfloat[\label{fig:delta_temp_lit}]{\includegraphics[width=0.76\linewidth]{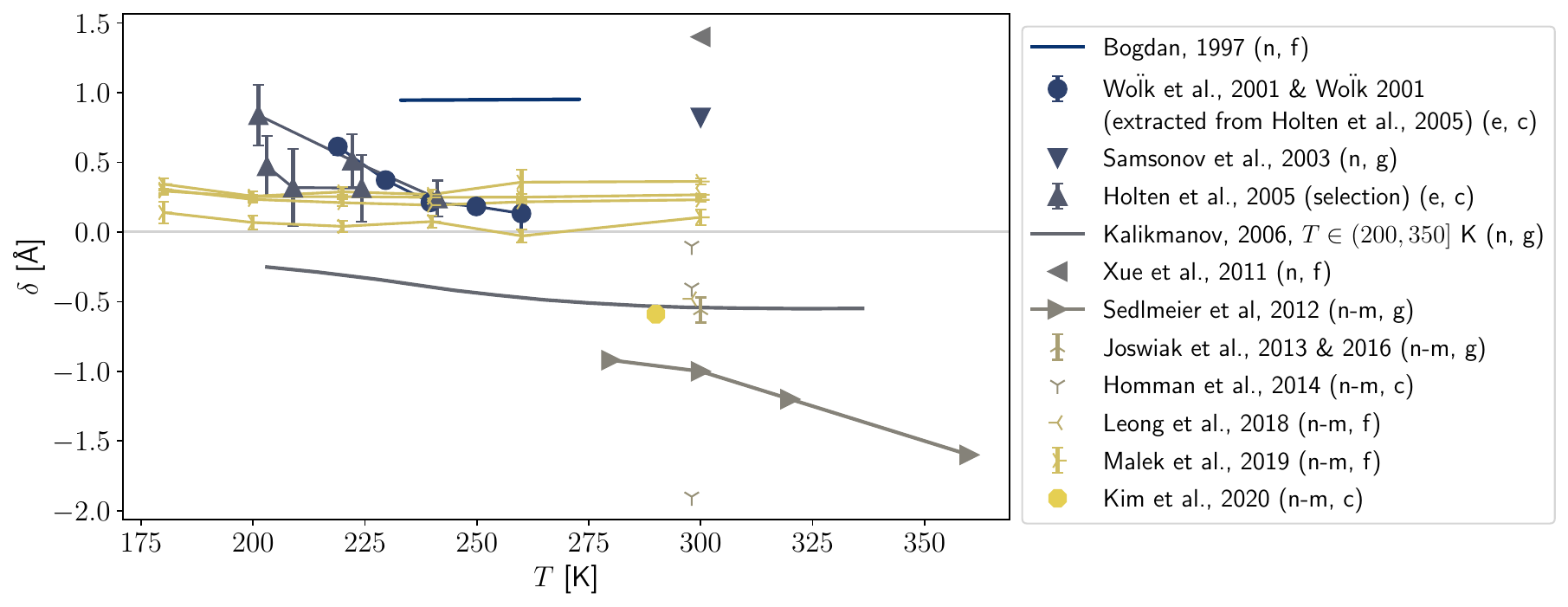}}%
    \subfloat[\label{fig:surf_tens_eqs_lit}]{\includegraphics[width=0.24\linewidth]{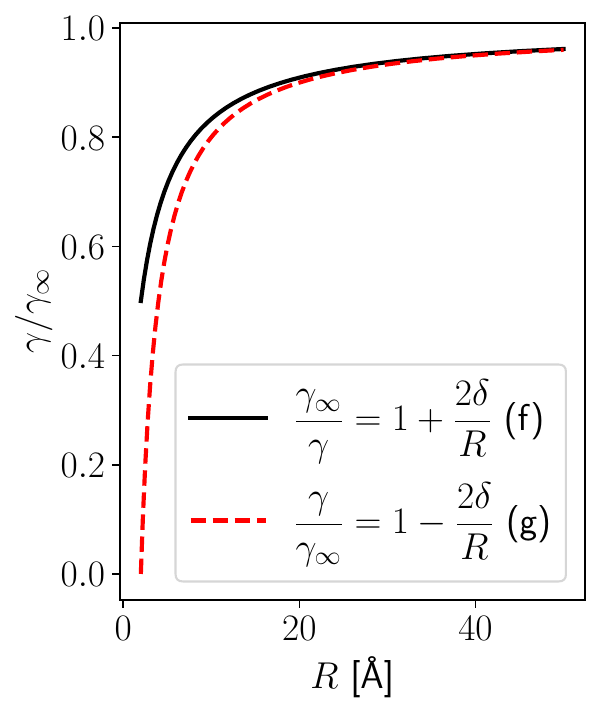}}
    \caption{(a) $\delta$ values with respect to temperature reported in the literature~\cite{Bogdan1997,Wolk2001,Wolk2001b,Samsonov2003,Holten2005,Kalikmanov2006,Xue2011,Sedlmeier2012,Joswiak2013,Joswiak2016,Homman2014,Leong2018,Malek2019,Kim2020}. The terms in the parentheses are: ``e'' for experimental studies; ``n'' for numerical studies; ``n-m'' for numerical studies at the atomistic level;  ``f'' and ``g'' for the Tolman equations, shown in \cref{fig:surf_tens_eqs_lit}, that were used to fit $\delta$;  ``c'' for studies where $\delta$ was computed directly from \cref{eq:delta_def}; (b) Ratio of the surface tensions of a curvature to the planar limit as predicted by \cref{eq:gamma_Tolman}-- label f-- and \cref{eq:gamma_Tolman_2}-- label g--.}   
\end{figure*}

Since the capacitance $\mathcal{C}$ of a sphere is
\begin{equation}
    \mathcal{C} = 4\pi \cvp R
\end{equation}
and the resistance  $\mathcal{R}$ of a cylinder is
\begin{equation}
    \mathcal{R} = \dfrac{ L}{\pi \xi R^2},
\end{equation} 
the characteristic time of the electric circuit shown in \cref{fig:grain_conductivity_scheme} is
\begin{equation}
    \tau_c = \mathcal{R} \mathcal{C} = \dfrac{4\cvp L}{\xi R}.
\end{equation}
Assuming that $4L/R$ is of order unity, the characteristic time is again  $\tau_c = \cvp / \xi$.

The conductivity of aqueous solutions (not related to dusty plasma conditions) arises mainly from the proton-hopping of \ch{H+} and \ch{OH-} ions, known as Grotthu{\ss} mechanism \cite{Marx2010}. Measurements vary significantly among experimental studies and have a strong temperature dependence \cite{Glen1975}. The conductivity of pure ice at $-40$ {\textcelsius} is $\xi=3\times 10^{-9}$ S/m. Using this conductivity value, $\tau_c\approx 3 \times 10^{-3}$ s. Therefore, it is reasonable to assume that ice grains behave as conductors on time scales of a fraction of a second or longer. Note that the above conductivity value was measured at a temperature that is significantly higher than the temperature of ice grains in astrophysical dusty plasmas \cite{Shukla2001,Woitke2009,Nicolov2024}.  At lower temperatures, the characteristic time $\tau_c$ is expected to be longer. However, it should be noted that the characteristic time scales in astrophysics are also much longer than $\tau_c$ (e.g., years to millions of years).

Even though solvated electrons are not free to move but rather are trapped in cavities, they can still diffuse among the water molecules. Solvated electrons diffuse in a different manner from the Grotthu{\ss} mechanism. Herbert \cite{Herbert2019} describes the diffusion of solvated electrons as a ``librationally-driven oozing'' mechanism. Specifically, the motion of water molecules around the solvated electron can cause the collapse of the cavity and the simultaneous creation of an adjacent one. The conductivity of solvated electrons at 25 {\textcelsius} is approximately 89\% of the conductivity of \ch{OH-} and half the conductivity of \ch{H+}.\cite{Hart1969} The conductivity of solvated electrons has a very strong temperature dependence. At 38 {\textcelsius} it is equal to the conductivity of hydroxide \cite{Schmidt1992}. Above this temperature, solvated electrons diffuse significantly faster than hydroxide, whereas below this temperature, there is a slight decrease in their mobility. This observation is in agreement with the ``librationally-driven oozing'' mechanism since lower temperatures mean a decrease in the kinetic energy of the water molecules and thus it is less likely for a cavity to collapse. The mobility of solvated electrons is comparable to the mobilities of other charged species. Therefore, the conductivity assumption for grains with solvated electrons is only valid in the macroscale. 

Consequently, \gls{oml} theory needs to be considered as an approximation because it assumes that grains are spherical when, in fact, they are not, and it assumes grains are conductors, when, in fact, the grain charge distribution may deviate from that of a conductor. Therefore, the use of the \gls{oml} theory needs to be justified based on the mobility of the charged species within the grain and the overall time scales under consideration.

\section{Surface tension of neutral and charged nanoclusters}
\label{sec:surf_tens}

This section investigates the surface tension of neutral and charged species. Attention is given to the surface tension of nanosized curvatures. According to Tolman~\cite{Tolman1949}, the surface tension $\gamma$ of a small liquid droplet is size-dependent and is described by
\begin{equation}
  \gamma = \dfrac{\gamma_{\infty}}{1+ \frac{2\delta}{r}},
  \label{eq:gamma_Tolman}
\end{equation}
where $\gamma_{\infty}$ is the surface tension of the interface in the planar limit and $r$ is the radius of the spherical droplet. The Tolman's length $\delta$ is defined as
\begin{equation}
\delta=R_e^{*}-R_s^{*},
\label{eq:delta_def}
\end{equation}
where $R_e^{*}$ and $R_s^{*}$ are respectively the radii of the equimolar and interface surfaces (for the definitions of these radii the reader is referred to Ref.~\citenum{Bogdan1997}).

\begin{figure*}[!ht]
    \centering
    \includegraphics[width=0.33\linewidth]{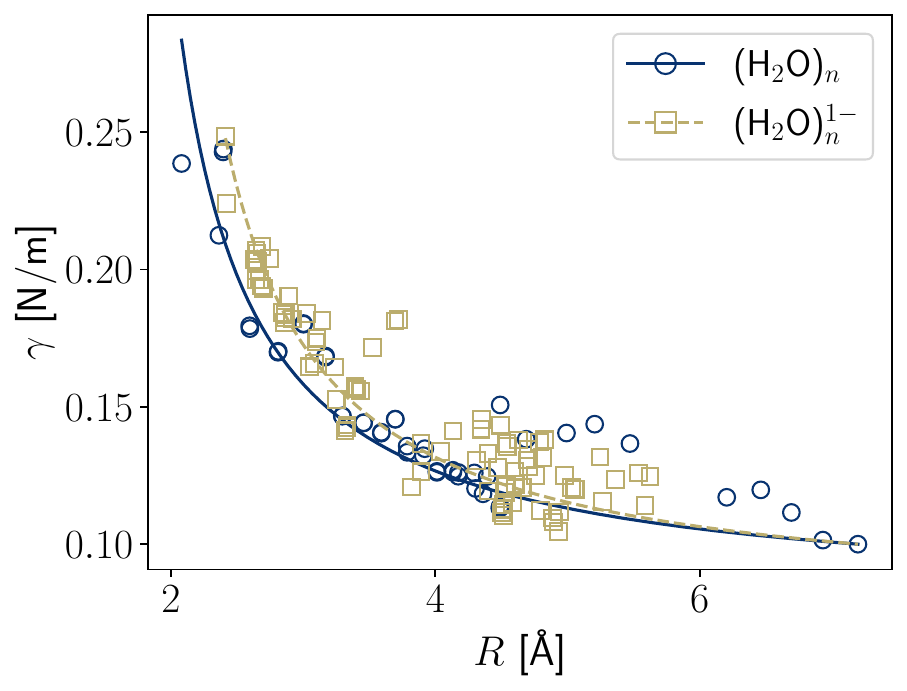}\hfill%
    \includegraphics[width=0.33\linewidth]{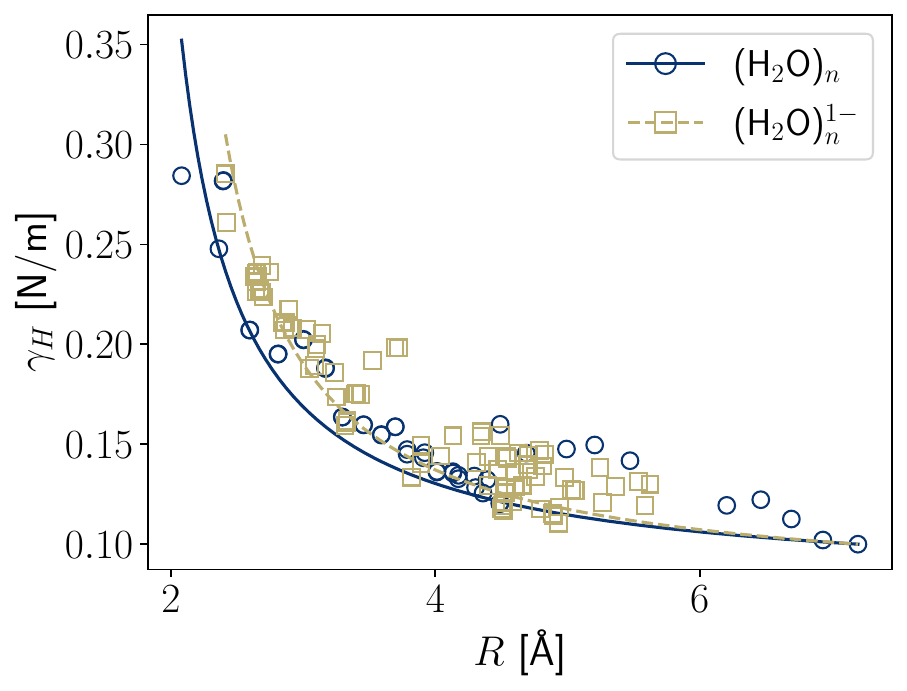}\hfill%
    \includegraphics[width=0.33\linewidth]{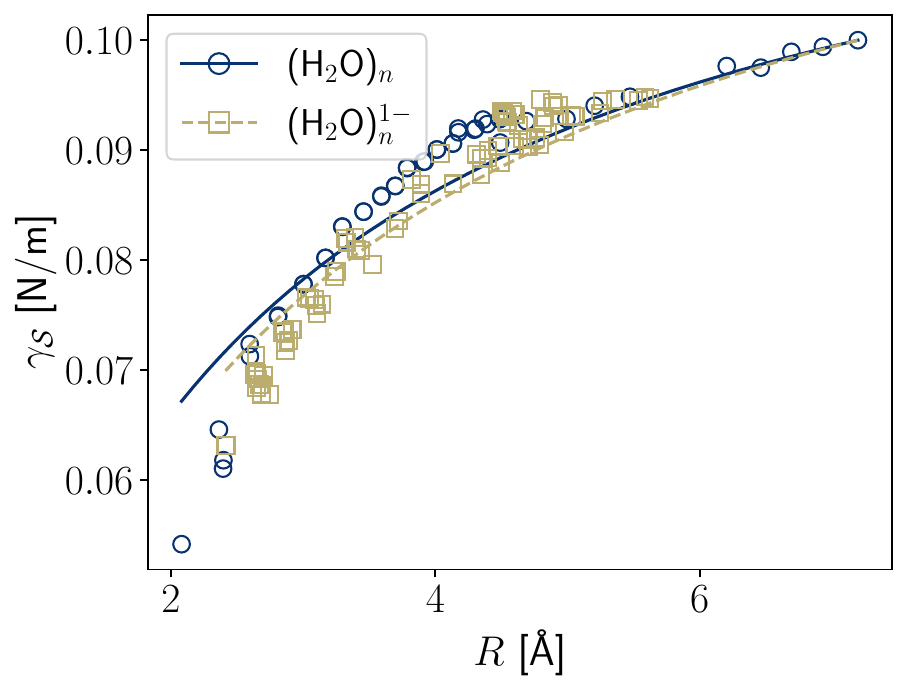}
    \caption{Total, enthalpic, and entropic surface tension of neutral and charged water clusters at $T=100$ K. It is assumed that $\gamma_{\infty \rightarrow (\mathrm{H}_2\mathrm{O})_{66}} = 0.1$ N/m. The lines correspond to fitted Tolman curves. The fitted $\delta$ values are shown in \cref{tab:delta_fitted}.}
    \label{fig:surface_tension}
\end{figure*}

Several studies apply the following expression for the surface tension
\begin{equation}
  \gamma = \gamma_{\infty}\left(1- \dfrac{2\delta}{r}\right),
  \label{eq:gamma_Tolman_2}
\end{equation}
which is a polynomial expansion of the curvature. These equations give slightly different surface tension curves for the same $\delta$ value as shown in \cref{fig:surf_tens_eqs_lit}. In both \cref{eq:gamma_Tolman,eq:gamma_Tolman_2}, the validity and significance of the neglected leading terms (such as the rigidity constant) have been debated in the literature \cite{Rowlinson1994,VanGiessen2002}. Moreover, contradictory findings have been presented in the literature on the dependence of Tolman's length $\delta$ on temperature, the state of matter of the interface (bulk material and ambient environment), and even on the equimolar radius. More importantly, the magnitude and even the sign of $\delta$ have been debated in the literature, even under the same conditions.

Values of $\delta$, obtained from several experimental and numerical studies, are shown in \cref{fig:delta_temp_lit}. These values span over a wide range, from approximately $-2$ {\AA} to 1.5 {\AA}. Bogdan \cite{Bogdan1997} obtained a positive $\delta$ using Gibbs adsorption, which slightly decreased with decreasing temperature. The experimental studies of Refs.~\citenum{Wolk2001,Wolk2001b,Holten2005} predict a positive and increasing Tolman length with decreasing temperature. However, Kalikmanov \cite{Kalikmanov2006} developed a model, termed Mean-field Kinetic Nucleation Theory (MKNT), and predicted negative $\delta$ values from the experimental results of \citenum{Wolk2001}.

Results are less consistent among numerical \gls{md} studies. \gls{md} studies use different approaches to compute the surface tension of small \ch{H2O} clusters. For instance, Joswiak et al. \cite{Joswiak2013,Joswiak2016} and Lau et al. \cite{Lau2015} use the mitosis method, according to which a single water cluster is gradually separated into two clusters by holding the molecules together via spring forces. Even though the same method was used, the studies predicted $\delta$ values with opposite signs, and the authors were unable to explain these discrepancies. Malek et al. \cite{Malek2019} computed Tolman's length from both the pressure tensor and the Laplace pressure and got an overall positive Tolman length. More recently, Kim and Jhe \cite{Kim2020} used the 2PT method \cite{Lin2010} to compute the Helmholtz free energy from \gls{md} simulations and obtained a negative Tolman length. This result was consistent with their \gls{dft} computations. 

Overall, the magnitude and sign of the Tolman length for water are not uniquely agreed upon in the literature. Atomistic methods directly model the interactions among water molecules and, in principle, should offer a better evaluation of the surface tension of small clusters compared to theoretical approximations or experimental interpretations. It is also noted that, as detailed below, $\delta$ depends heavily on the interatomic potential. \gls{dft} offers a more detailed description of molecular interactions compared to \gls{md}. Therefore, it is expected that the surface tension of small curvatures is predicted more accurately with \gls{dft} computations. In this study, \gls{dft} is used to compute the surface tension of small water clusters. Specifically, the surface tension of water is computed through the thermochemical analysis of relaxed geometries at different temperatures. In addition, the impact of charges on the surface tension of water clusters is investigated. 

Surface tension is related to Gibbs free energy $G = H - T \mathcal{S}$ as
\begin{equation}
\begin{aligned}
    \gamma &= \dfrac{\Delta G}{A} = \dfrac{\Delta H - T \Delta \mathcal{S}}{A}\\
           &=\dfrac{\left(H_{r} - \frac{N_{r}}{N_{p}} H_{p} \right) - T \left(\mathcal{S}_{r} - \frac{N_{r}}{N_{p}} \mathcal{S}_{p}\right)}{A},
    \label{eq:gamma_atomistic}
\end{aligned}
\end{equation}
where $N$ and $A$ denote the number of molecules and the surface area of the cluster. The subscripts $r$ and $p$ denote the small spherical cluster and planar limit, respectively. The term $\Delta H/A$ is the enthalpic contribution to the surface tension, $\gamma_H$, whereas the term $- T \Delta \mathcal{S}/A$ is the entropic contribution, $\gamma_\mathcal{S}$. Entropy $\mathcal{S}$ is given by
\begin{equation}
\begin{aligned}
    \mathcal{S} = \mathcal{S}_{el}+\mathcal{S}_{vib}+\mathcal{S}_{rot}+\mathcal{S}_{trans}.
    \label{eq:dft_entropy}
\end{aligned}
\end{equation}
The $\mathcal{S}$ terms in the \gls{rhs} of \cref{eq:dft_entropy} are respectively the electronic, vibrational, rotational, and translational entropies. The enthalpy $H$ is
\begin{equation}
\begin{aligned}
    H = U + k_B T,
\end{aligned}
\end{equation}
where $U$ is the inner enthalpy. $U$ is computed using
\begin{equation}
\begin{aligned}
    U = E_{el} + E_{ZPE} + E_{vib} + E_{rot} + E_{trans}.
\end{aligned}
\end{equation}
$E_{el}$ is the total energy from the electronic structure calculation; $E_{ZPE}$ is the zero temperature vibrational energy; $E_{vib}$ is the finite temperature correction to $E_{ZPE}$ due to population; $E_{rot}$ and $E_{trans}$ are respectively the rotational and translational thermal energies. It should be noted that \cref{eq:gamma_atomistic} is based on thermodynamic equilibrium arguments. However, the plasma and grains should not be in thermodynamic equilibrium at the timescales of interest to the previous sections, i.e., at timescales comparable to the deformation time of grains or to the mean collision time of grains and plasma species.

The above parameters are computed from the vibrational modes of relaxed geometries as detailed in Refs. \citenum{Grimme2012,Herzberg1945} and using the ORCA software \cite{ORCA}. \gls{dft} simulations are conducted using the B3LYP-D3/6-311++G** level of theory. The thermochemistry is analyzed at 100, 273, and 298 K. Simulation details and geometries are detailed in \cref{sec:si_surf_tens}. Since calculating the planar limit of ice, $\gamma_\infty$, is computationally unfeasible using \gls{dft}, a value of $\gamma_{\infty}=0.1$ N/m is assumed. Therefore, in \cref{eq:gamma_atomistic}, all surface tension values are given with respect to 0.1 N/m.

\begin{figure}[!t]
    \centering
    \includegraphics[width=0.66\linewidth]{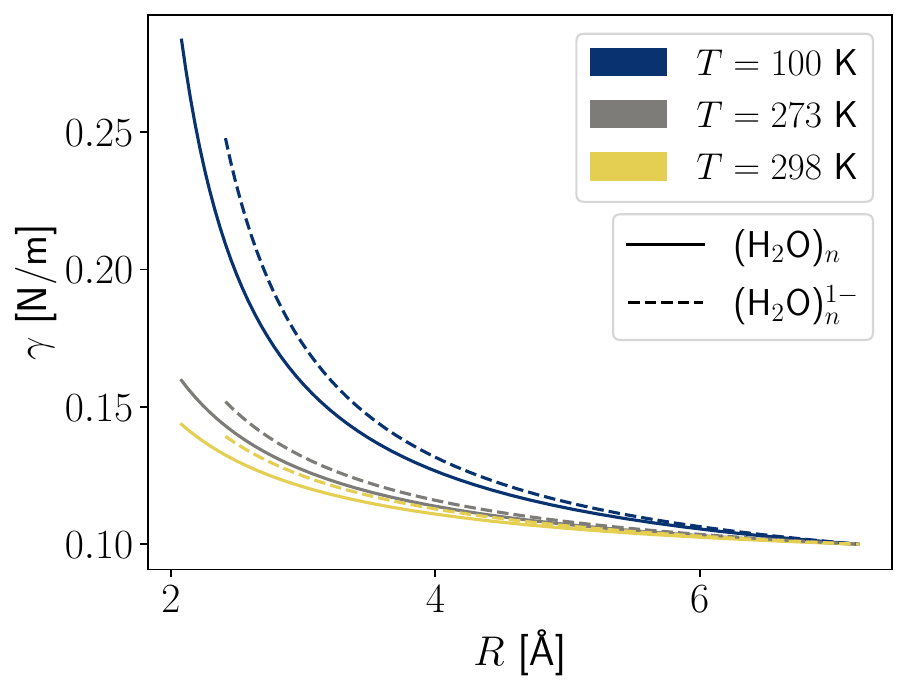}
    \caption{Total surface tension of neutral and charged water clusters at different temperatures. The lines correspond to Tolman curves, fitted to the \gls{qm} computations. It is assumed that $\gamma_{\infty \rightarrow (\mathrm{H}_2\mathrm{O})_{66}} = 0.1$ N/m. The fitted $\delta$ values are shown in \cref{tab:delta_fitted}.}
    \label{fig:surface_tension_temp}
\end{figure}

\begin{table}[!ht]
    \centering
    \begin{tabular}{c|cc}
         & \multicolumn{2}{c}{$\delta$ [{\AA}]} \\
       $T$ [K]  & $(\ch{H2O})_n$  & $(\ch{H2O})_n^{1-}$ \\
       \hline
       100  & $-0.75$ & $-0.83$ \\
       273  & $-0.47$ & $-0.53$ \\
       298  & $-0.40$ & $-0.45$ \\
       \hline
    \end{tabular}
    \caption{Fitted $\delta$ values from \cref{fig:surface_tension_temp}.}
    \label{tab:delta_fitted}
\end{table}

\Cref{fig:surface_tension} shows the surface tension obtained from different geometries and the fitted Tolman curves. The radius of the clusters is computed assuming a spherical volume. The area of the clusters is computed according to the Connolly surface using the Molovol software \cite{Maglic2022}. In addition, the enthalpic and entropic contributions of the surface tension are reported. The temperature dependence of the $\gamma$ terms is shown in \cref{fig:surface_tension_temp}. The values of the fitted Tolman lengths $\delta$ are given in 
\cref{tab:delta_fitted}. Results show that for all conditions, the surface tension of neutral clusters increases as the size of the cluster decreases. This result agrees with the \gls{dft} studies of Refs.~\citenum{Barrett1999,Kim2020}. As the temperature of the system increases, the surface tension decreases. This is attributed to the $-T$ term of \cref{eq:gamma_atomistic}.

Joswiak et al. \cite{Joswiak2016} explained the size dependence of the surface tension by means of the enthalpic and entropic contributions. Since these terms are of the same order of magnitude, they contribute equally to the free energy. The competition between the enthalpic and entropic contributions determines the behavior of the surface tension with respect to the size of the droplet. Enthalpy depends on the strength and number of \ch{H2O}--\ch{H2O} bonds. Therefore, as the size of the droplet decreases, there are more broken bonds per surface area, leading to an increase in surface energy. Entropy is related to the correlations of the molecules within the droplets. As the size of the droplet decreases, more molecules are correlated with each other, i.e., $\zeta/R$ increases, where $\zeta$ is the correlation length \cite{Joswiak2016}. Thus, entropy decreases as the size of the droplet decreases.

In our study (cf. \cref{fig:surface_tension}), $\gamma_H$ increases faster than $\gamma_\mathcal{S}$ decreases with decreasing size of the cluster. As a consequence, the surface tension increases with decreasing size. It was also observed that the surface tension of the geometries generated in this study by cutting a large bulk (cf. \cref{sec:si_surf_tens}) exhibit a higher surface tension compared to the more organized geometries obtained from the literature. The reason is that in our geometries, there are more \ch{H2O}--\ch{H2O} broken bonds than the structures from the literature and thus $\gamma_H$ is larger.

Electrons further increase the surface tension of small water clusters. As is evident from \cref{fig:surface_tension}, the magnitude of $\gamma_H$ increases when electrons are present. $\gamma_\mathcal{S}$ also increases but less significantly. These observations can be explained using the above theory. Solvated electrons cause several water-water bonds to break, which increases the enthalpic contribution. In addition, solvated electrons slightly disturb the geometry by reducing $\zeta/R$ and thus increasing the randomness in the clusters. Consequently, entropy increases when electrons are present (cf. \cref{fig:entropy_area}), and thus the entropic contribution $\gamma_\mathcal{S}$ decreases.

\section{Conclusions}

This study investigates the impact of electrostatic stresses on the structure of water grains inside a weakly ionized plasma. The theoretical analysis is validated using \glsfirst{md} simulations. In addition, the conductivity assumption of ice grains is assessed. Finally, the size dependence of the surface tension of neutral and charged water clusters is investigated.

Our findings indicate that electrostatic stresses can affect nanosized grains depending on the \textcolor{black}{floating} potential, surface tension of the material, and local radius of curvature. When electrostatic stresses dominate, the grain elongates. Our atomistic simulations predict that the surface tension of neutral and charged water clusters increases with decreasing \textcolor{black}{radius of} curvature. This surface tension increase may cause grains to \textcolor{black}{cease} elongating. Electrostatic stresses do not affect micro- and macrosized grains with small aspect ratios. The reason is that interelectron repulsions are weak on large grains since there are not enough electrons per unit surface area.

Our MD simulations of spherical grains charged to different potentials showed excellent agreement with the theoretical findings. In addition, our theoretical analysis showed that water-based grains behave as conductors on macro timescales. However, our \gls{qm} computations revealed that the strong binding nature of solvated electrons overcomes electron-electron repulsions. While water grains cannot be considered as conducting materials on nano timescales, the lifetime of astrophysical grains is much longer (seconds to millions of years), so they eventually behave as conductors.

\textcolor{black}{We also} identified timescales of the various phenomena associated with nanosized water grains in a weakly ionized plasma. Specifically, the following phenomena occur at different timescales \textcolor{black}{in} increasing order of magnitude (fastest to slowest): (i) grain deformation at a constant charge, (ii) grain-electron and grain-ion collisions, (iii) grain deformation at a constant potential, (iv) grains reaching a conductive behavior, (v) thermodynamic equilibrium of macroscale parameters.

\section{Supplementary Material}

The \glsfirst{si} contains numerical methods, simulation details, and additional calculations and graphs.

\begin{acknowledgments}
The authors would like to thank Prof. A. van Duin for discussions on solvated electrons and A. Nicolov for discussions on the Caltech dusty plasma experiment. The computations presented here were conducted using the Resnick High-Performance Computing Center, a facility supported by the Resnick Sustainability Institute at the California Institute of Technology. WAG received support from NSF CBET Award Number 2311117. PMB received support from NSF Award Number 2308558.
\end{acknowledgments}

%

\end{document}



\title{Can electrostatic stresses affect charged water structures in weakly ionized plasmas? -- Supporting Information}

\author{Efstratios M. Kritikos}
\affiliation{Department of Applied Physics and Materials Science, California Institute of Technology, Pasadena, 91125, United States}
\email{emk@caltech.edu}
 
\author{William A. Goddard III}
\affiliation{Department of Applied Physics and Materials Science, California Institute of Technology, Pasadena, 91125, United States}

\author{Paul M. Bellan}
\affiliation{Department of Applied Physics and Materials Science, California Institute of Technology, Pasadena, 91125, United States}

\date{\today}

\begin{abstract}
This document contains information in support of the manuscript.
\end{abstract}

\maketitle

\section{Electrostatic and surface tension stresses}
\label{sec:si_elec_surf_stress}

\begin{figure}[!ht]
    \centering
    \subfloat[$a=10$ nm]{\includegraphics[width=0.5\linewidth]{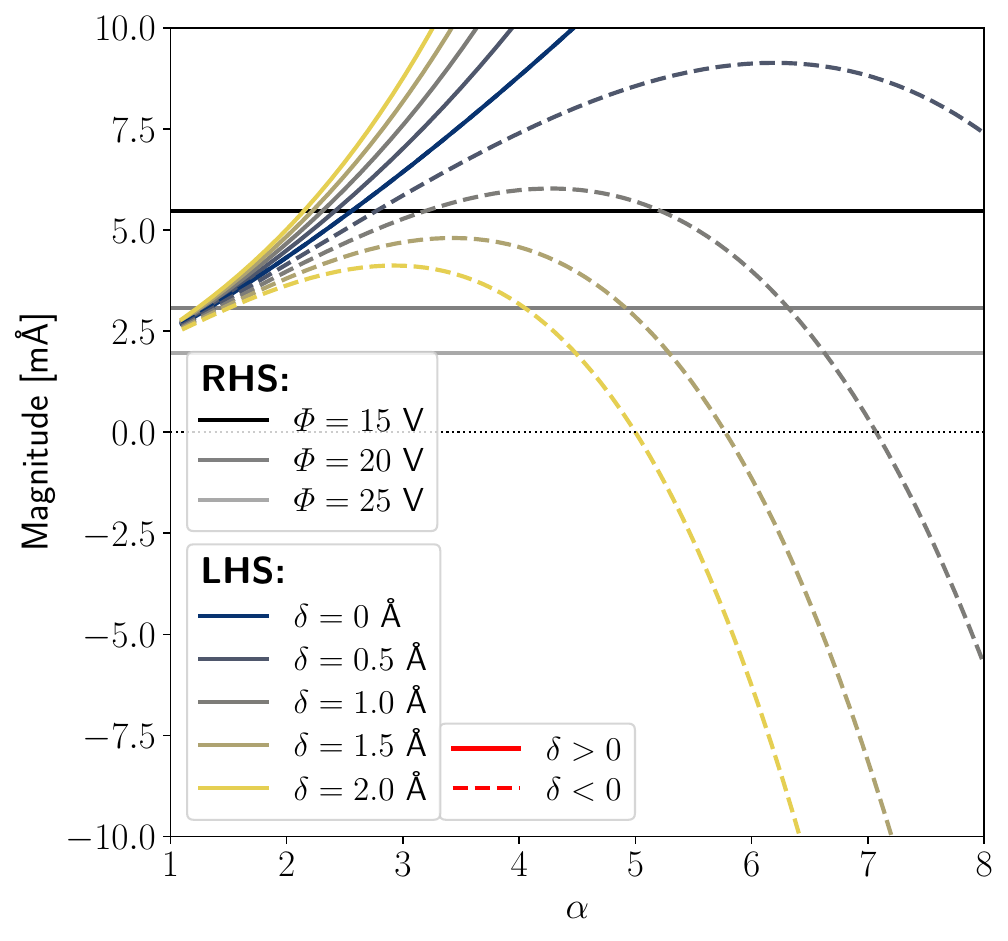}}%
    \subfloat[$a=1000$ nm]{\includegraphics[width=0.5\linewidth]{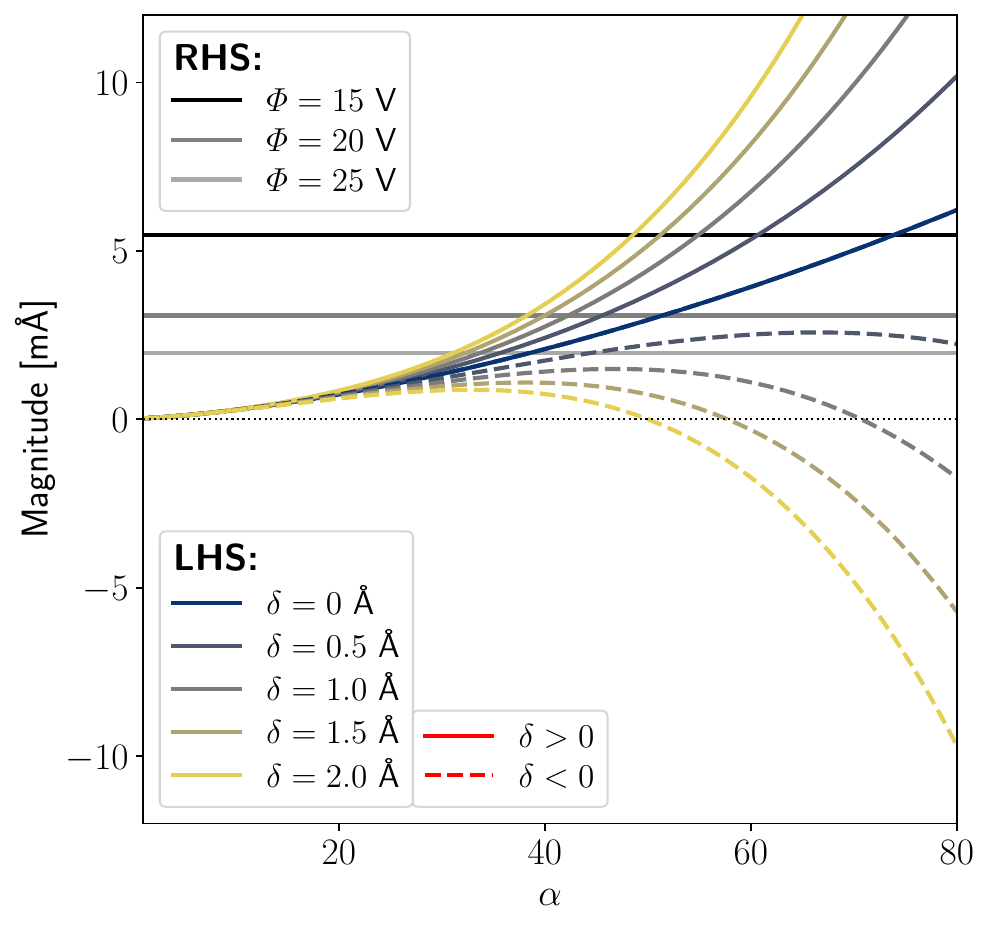}}
    \caption{Plot of the \gls{rhs} and the \gls{lhs} of \cref{eq:est_ineq} of the manuscript for $\gamma_{\infty}=0.109$ N/m. Solid lines represent positive $\delta$ values, while negative values are plotted using dashed lines.}
    \label{fig:est}
\end{figure}

\begin{figure}[!ht]
    \centering
    \subfloat[$\gamma_{\infty}=0.109$ N/m]{\includegraphics[height=7cm]{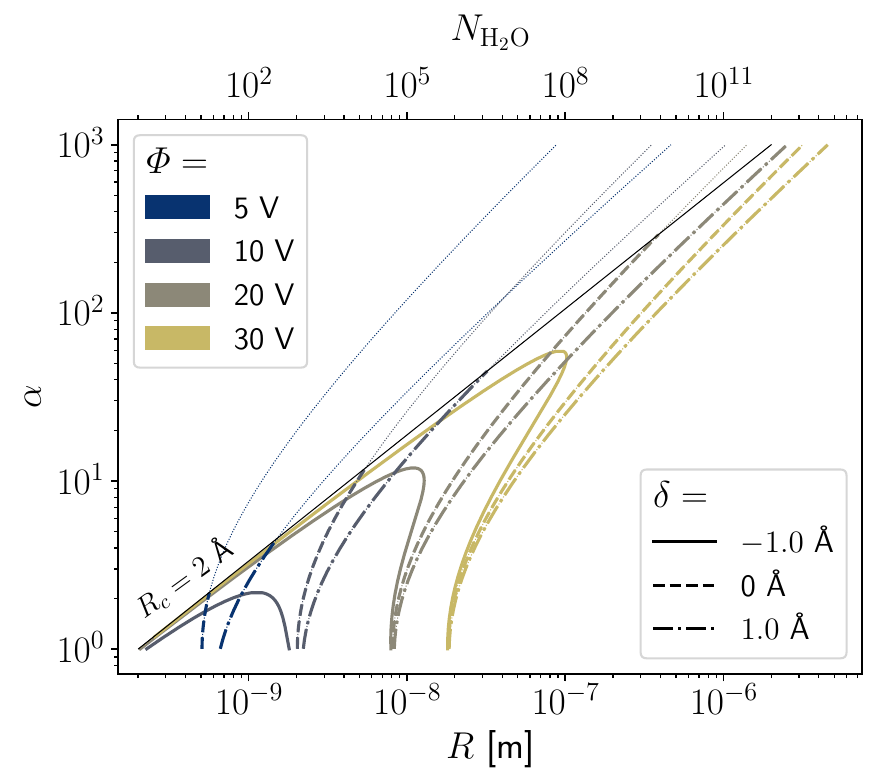}}\hspace{10pt}%
    \subfloat[$\delta=0$ {\AA}]{\includegraphics[height=7cm]{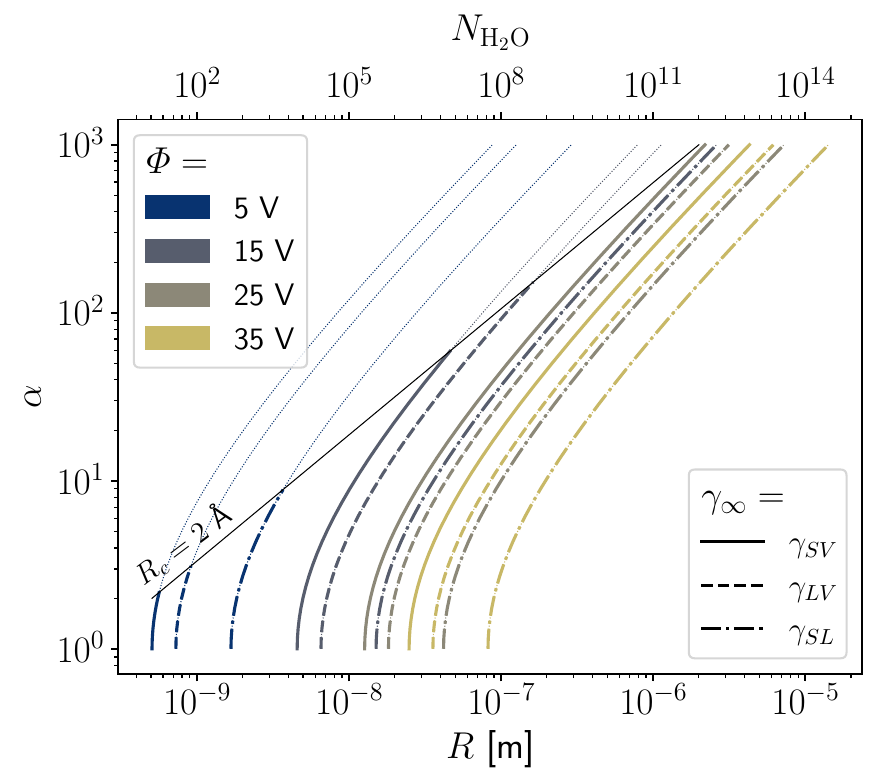}}
    \caption{Plot of \cref{eq:est_ineq_solR} of the manuscript for different material and plasma properties. The area to the left of each $R$ solution corresponds to $\tau_{\mathcal{E}} > \tau_{\gamma}$, whereas the area to the right corresponds to $\tau_{\mathcal{E}} < \tau_{\gamma}$. $R_c=2$ {\AA} is set as a curvature size limit for water clusters. Dotted lines correspond to solutions with an $R_c$ value lower than the 2 {\AA} limit.}
    \label{fig:est_sol_large_manylines}
\end{figure}

\Cref{fig:est} shows the \gls{rhs} and the \gls{lhs} of \cref{eq:est_ineq} of the manuscript. In addition, solutions of \cref{eq:est_ineq_solR} are plotted in \cref{fig:est_sol_large_manylines}. Results show that smaller grain sizes, lower surface tensions, and higher electric potentials aid electrostatic stresses to overcome surface tension stresses, causing the grain to become elongated. Negative Tolman lengths $\delta$ lead to an increase in the surface tension with decreasing curvature size. Therefore, a negative $\delta$ may halt the deformation of the grain. On the contrary, a positive $\delta$ decreases the surface tension and thus aids the deformation.

\pagebreak

\Cref{eq:est_ineq} is solved analytically with respect to $\alpha$. We consider large grains for which Tolman's correction to the surface tension can be neglected (i.e., $a>>\delta$) or materials with $\delta=0$. In that case, the equality of \cref{eq:est_ineq} is
\begin{equation}
R = \dfrac{\cvp \varPhi^{2}\left(\alpha^{2}-1\right)}{\gamma_{\infty} \alpha^{2/3}\left[\ln\left(\frac{1+\sqrt{1-1/\alpha^{2}}}{1-\sqrt{1-1/\alpha^{2}}}\right)\right]^{2}}.
\label{eq:est_sol_large_R}
\end{equation}

In addition, solutions for which $\alpha^2 >> 1$ are considered. Therefore, the natural logarithm term in \cref{eq:est_ineq} is simplified using the Taylor expansion $\sqrt{1-1/\alpha^2} \approx 1 - 1/(2 \alpha^2)$, and becomes $\ln(4\alpha^2)$. For simplicity, the RHS of \cref{eq:est_ineq} is substituted by the constant $C=\gamma_{\infty}/\cvp \varPhi^{2}$. Moreover, we write \cref{eq:est_ineq} with respect to $a$ and $\alpha$ and solve analytically. Therefore, the equation becomes
\begin{equation}
\begin{aligned}
\alpha^2- a C\ln(4\alpha^2) - 1 &= 0, \\
\ln(4\alpha^2) &= \sqrt{\dfrac{\alpha^2-1}{aC}}, \\
\ln(4\alpha^2) &\approx \dfrac{\alpha}{\sqrt{aC}},
\label{eq:est_ineq_large_2}
\end{aligned}
\end{equation}
where the final equation was simplified by the aforementioned assumption that $\alpha^2 >> 1$. \Cref{eq:est_ineq_large_2} can be reorganized in the form
\begin{equation}
\begin{aligned}
-\dfrac{\alpha}{2\sqrt{aC}} \exp(-\dfrac{\alpha}{2\sqrt{aC}}) = -\dfrac{1}{4\sqrt{aC}}.
\label{eq:est_ineq_large_3}
\end{aligned}
\end{equation}

\Cref{eq:est_ineq_large_3} has the form $y\mathrm{e}^y=x$ and thus can be solved using Lambert's $W$ functions \cite{Valluri2000} for $x \geq -1/\mathrm{e}$. Since both $y$, $x\in \mathbb{R}$, and $-1/\mathrm{e} \leq x < 0$, the solutions are of the form $y=W_0(x)$ and $y=W_{-1}(x)$. Therefore, the two solutions $\alpha_1$ and $\alpha_2$ of \cref{eq:est_ineq_large_3} are
\begin{equation}
\begin{aligned}
\alpha_1&=-2 \sqrt{aC}\ W_0{\left[-\dfrac{1}{4\sqrt{aC}}\right]},\\
\alpha_2&=-2 \sqrt{aC}\ W_{-1}{\left[-\dfrac{1}{4\sqrt{aC}}\right]}.
\label{eq:est_sol_alpha_large}
\end{aligned}
\end{equation}
The solution $\alpha_1$ is disregarded since it gives $\alpha$ solutions lower than 1, which contradicts the initial assumption. The solution $\alpha_2$ is valid in the range $-1/\mathrm{e} \leq -1/(4\sqrt{aC}) < 0$. The right inequality is always satisfied, whereas the left inequality is satisfied when
\begin{equation}
\begin{aligned}
\dfrac{\gamma_{\infty} a}{\varPhi^2} \ge \dfrac{\cvp \mathrm{e}^2}{16}.
\label{eq:est_sol_alpha_large_final}
\end{aligned}
\end{equation}

\begin{figure}[!ht]
    \centering
    \includegraphics[width=0.5\linewidth]{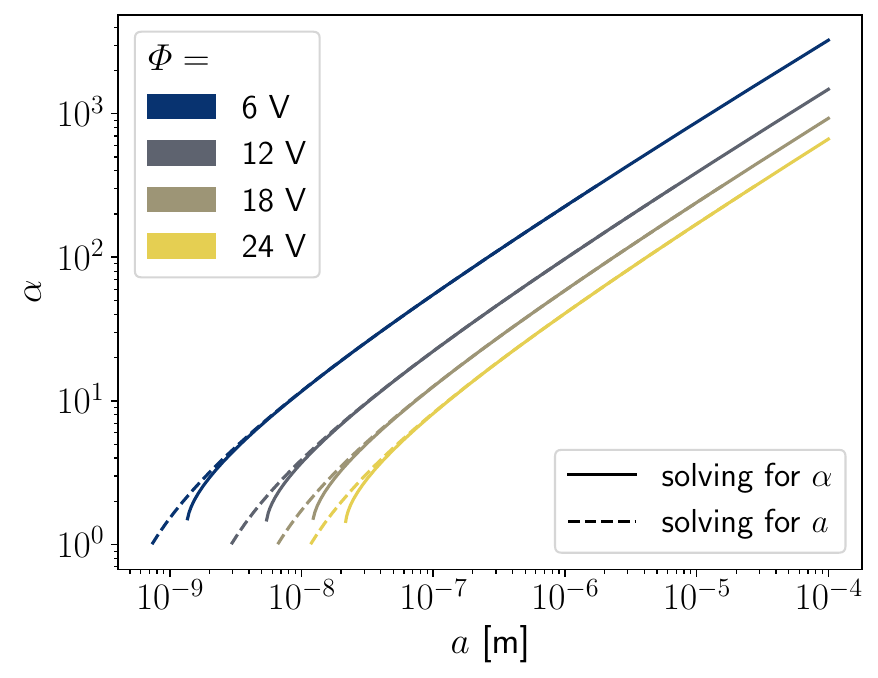}
    \caption{Comparison of \cref{eq:est_sol_large_R} and \cref{eq:est_sol_alpha_large_final}.}
    \label{fig:est_sol_alpha_large}
\end{figure}

In conclusion, the solution of \cref{eq:est_ineq} of the manuscript with respect to $\alpha$ is equal to
\begin{equation}
    \alpha=-2 \sqrt{\dfrac{a\gamma_{\infty}}{\cvp \varPhi^{2}}}\ W_{-1}{\left[-\dfrac{1}{4}\sqrt{\dfrac{\cvp \varPhi^{2}}{a\gamma_{\infty}}}\right]},
\end{equation}
under the conditions that $\alpha^2>>1$ and $\gamma_{\infty} a/\varPhi^2 \ge \cvp \mathrm{e}^2/16$. The validation of this equation is shown in \cref{fig:est_sol_alpha_large}.

\section{Validation using molecular dynamics}
\label{sec:si_valid_md}

\begin{table}[!htb]
    \centering
    \begin{tabular}{cc|cc}
        $\gamma_{\infty}$ [N/m] & $\delta$ [{\AA}] & $\varPhi_{sph,lim}$ [V] & $Q_{sph,lim}$ [e]\\
        \hline
        $0.076$ & $-0.5$ & $9.46$ & $-16.42$ \\
        $0.076$ &  $0.0$ & $9.26$ & $-16.08$ \\
        $0.109$ & $-0.5$ & $11.32$ & $-19.66$ \\
        $0.109$ &  $0.0$ & $11.10$ & $-19.26$ \\
         \hline
    \end{tabular}
    \caption{Potential and charge of spherical grains that satisfy the equality of \cref{eq:est_ineq} of the manuscript under different surface tension and Tolman length values.}
    \label{tab:phi_q_limit}
\end{table}

\begin{figure}[!ht]
    \centering
    \subfloat[$\delta=-0.5$ {\AA} and $\varPhi=\varPhi_{sph,lim}=9.46$ V.]{\includegraphics[width=0.5\linewidth]{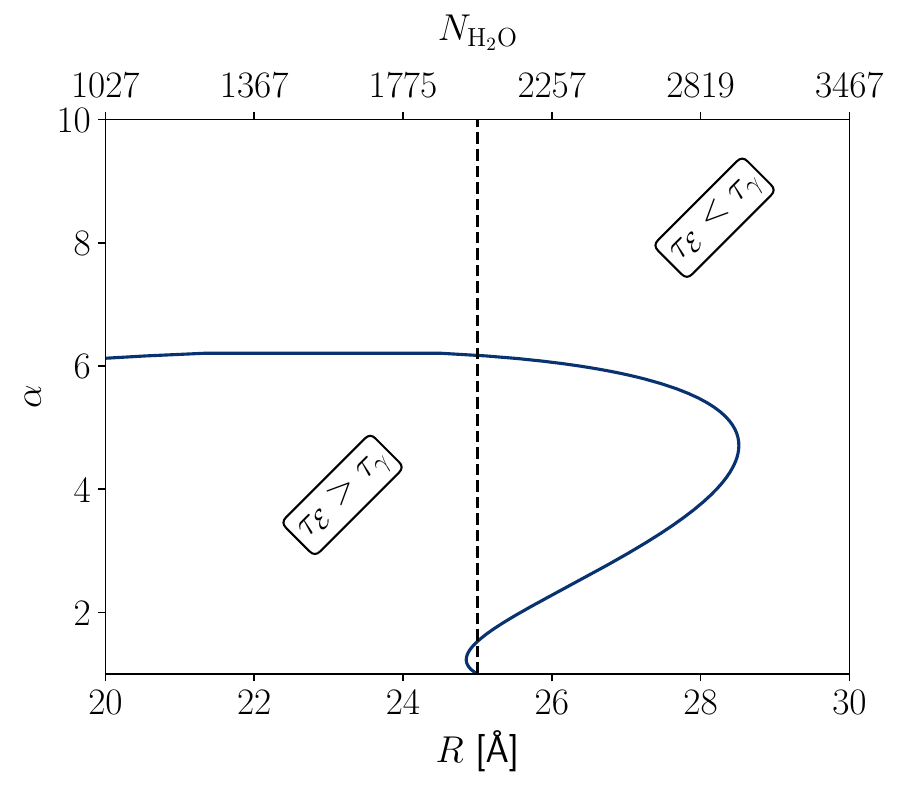}}%
    \subfloat[$\delta=0$ {\AA} and $\varPhi=\varPhi_{sph,lim}=9.26$ V.]{\includegraphics[width=0.5\linewidth]{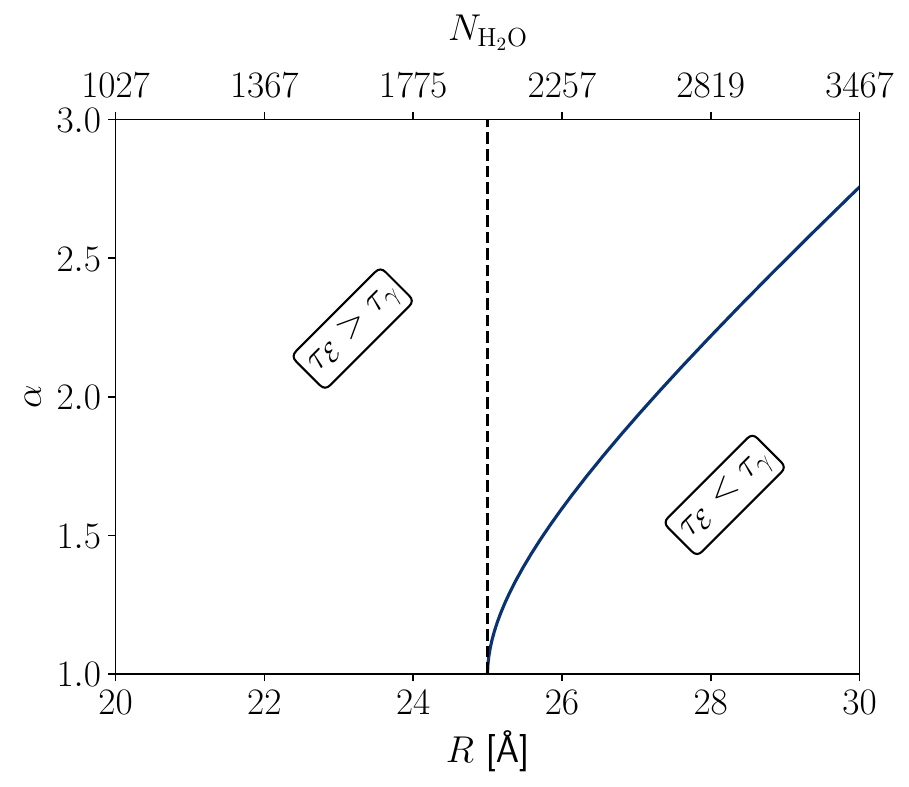}}
        
    \caption{Plot of \cref{eq:est_ineq_solR}, which shows when electrostatic stresses will overcome the surface tension stresses when $R=25$ {\AA}. Results are plotted for water with $\gamma_{\infty}=0.076$ N/m and two different $\delta$ values.}
    \label{fig:est_sol_large_delta_md_case}
\end{figure}

The necessary electric potential and equivalent charge to affect the structure of the grains discussed in \cref{sec:md_valid} are shown in \cref{tab:phi_q_limit}. In addition, solutions of \cref{eq:est_ineq_solR} are plotted in \cref{fig:est_sol_large_delta_md_case} for liquid water grains.

\begin{figure}[!ht]
    \centering
    \includegraphics[width=0.5\linewidth]{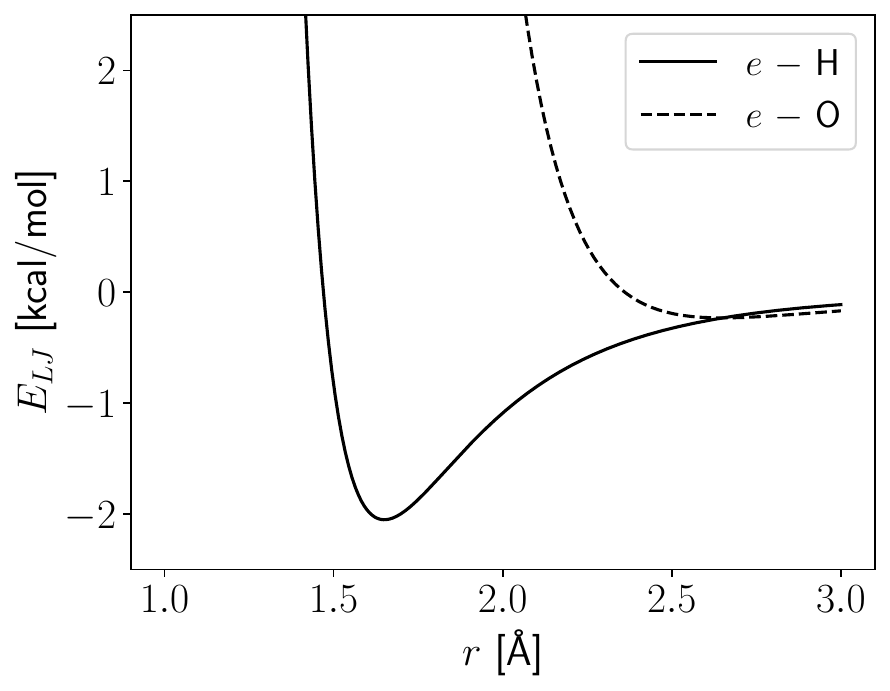}
    \caption{\gls{lj} potential for the \ch{$e$-H} and \ch{$e$-O} interactions.}
    \label{fig:pot_lj}
\end{figure}

The interactions between electrons and hydrogen or oxygen atoms are shown in \cref{fig:pot_lj}. The \gls{lj} parameters for these curves are given in \cref{tab:lj_values} of the manuscript. These parameters were computed based on Fig. 6 of the study from Kumar et al. \cite{Kumar2015}. Kumar et al. computed the relative total energy of a structure with 1 solvated electron and 4 water molecules in a tetrahedral symmetry with respect to the distance between the \gls{csdf} and the nearest H atom. An implicit solvation scheme was used in their simulations. The relative energy was computed for different levels of theory.

Based on Fig. 6 from Kumar et al. \cite{Kumar2015}, the equilibrium \ch{$e$-H} distance is $r_{min}=1.65$ {\AA}. Since the SPC/E model has an \ch{O-H} distance of 1.0 {\AA} and the angle between \ch{$e$-H-O} should be 180$^{\circ}$, the \ch{$e$-O} distance is set to 2.65 {\AA}. The relation between $\sigma_{LJ}$ and $r$ is
\begin{equation}
    \sigma=\dfrac{r}{2^{1/6}}.
\end{equation}
Therefore, the $\sigma_{LJ}$ parameter is equal to 1.46998 {\AA} and 2.36088 {\AA} for the \ch{$e$-H} and \ch{$e$-O} pairs, respectively. 

Moreover, Fig.~6 from Kumar et al. \cite{Kumar2015} shows the energy difference between the equilibrium point and the \gls{csdf}--H distance of 2.1 {\AA}. Based on this relative energy, an $\epsilon_{LJ}$ value of 2.052 kcal/mol was fitted to the CCSD(T)/aug-cc-pvTz. This fitting approach does not match completely the $e$--H interactions since the relative energy of the system includes different types of interactions (Coulomb, van der Waals interactions) between electrons and water molecules and also interactions among different water molecules. However, as will be shown below, this approach reproduces with good accuracy the distances between electrons and H or O atoms. A more comprehensive calibration of this system will be done in future work. The $\epsilon_{LJ}$ value for the \ch{$e$-O} interaction is found to be 0.232 kcal/mol. This value is found by fitting the relative energy of Fig.~6 from Kumar et al. \cite{Kumar2015} at $r=2.65$ {\AA}.

The $e$--H and $e$--O distances are monitored to examine whether the solvated $e$ parameters are accurate. These distances are computed for the case with $N_e=16$ shown in \cref{fig:md_results_220K}. The distances are averaged over all electrons and the last 3 trajectories. The average $e$--H distance was 2.04 {\AA} and the average $e$--O distance was 2.56 {\AA}. For the left solvated electron in the cluster of \cref{fig:n24_21_24_spindens}, the average distances between the \gls{csdf} and the nearest H and O atoms were 1.97 and 2.87 {\AA}, respectively. Therefore, the fitted LJ parameters can predict with good accuracy the cavity size of a solvated electron.

Spherical grains with a radius of 25 {\AA} are created by randomly placing water molecules in a sphere using the Packmol software \cite{Martinez2009}. A minimum distance of 2 {\AA} is kept between adjacent water molecules. The total number of water molecules is 2181. Different numbers of electrons are placed on a sphere with a radius of 20 {\AA} being concentric with the grain. To emulate the charge distribution of a conductive sphere, in which the electrons are uniformly placed on the surface, the Fibonacci sphere algorithm was used \cite{Marques2013}. According to this algorithm the $x$, $y$, and $z$ Cartesian coordinates of a particle $k$ are
\begin{equation}
\begin{aligned}
    x_k&=R\cos(k\pi \left(3 - \sqrt{5}\right))\sqrt{1-\left(\dfrac{2k+1}{N} - 1\right)^2},\\
    y_k &= R\left(\dfrac{2k+1}{N} - 1\right),\\
    z_k&=R\sin(k\pi \left(3 - \sqrt{5}\right))\sqrt{1-\left(\dfrac{2k+1}{N} - 1\right)^2},\\
\end{aligned}
\end{equation}
where $N$ is the number of all particles and $R$ is the radius of the sphere.

All simulations are performed using the LAMMPS software~\cite{Plimpton1995}. A cutoff distance of 10 {\AA} and 25 {\AA} is used for the \gls{lj} and Coulomb interactions, respectively. Initially, all geometries undergo an energy minimization process. Then, initial atomic velocities are assigned based on an MB distribution at a temperature of 10 K. The grain is gradually heated to the target temperature for a total duration of 40 ps. The equations of motion are discretized and solved using the Verlet algorithm with a timestep of 2~fs. The results of \cref{fig:md_results_220K,fig:md_results_273K} are collected from simulations performed at a fixed temperature. In all simulations, the NVT ensemble is applied. The temperature of the system is regulated with the Nos\'e-Hoover thermostat~\cite{Nose1984,Hoover1985} on the translational degrees of freedom of atoms. The damping constant of the thermostat is 100 fs.

\section{Solvated electrons and conductivity}
\label{sec:si_solv_cond}

\gls{qm} simulations were performed using the ORCA software \cite{ORCA}. The \gls{vebe} results in \cref{fig:cond_vebe} of the manuscript are computed for geometries reported in the literature or built in this study. Specifically, a total of 64 geometries are taken from Refs. \citenum{Zhan2003,Williams2008,Herbert2005,Kumar2015}. In addition, two geometries for $n=8$ and $16$ were generated by cutting an initial bulk amorphous ice structure. The bulk amorphous structure was generated using the Genice2 software \cite{Matsumoto2017}.

For the reported \gls{vebe} values, an initial geometry optimization was performed on all geometries using \gls{dft}. The B3LYP exchange-and-correlation (XC) functional is used with the D3 dispersion correction and the 6-311++G** basis set. The convergence criteria are shown in \cref{tab:go_conv_crit}. \glspl{vebe} were computed using \gls{mp2} computations with the 6-311++G** basis set. Only 64 of all geometries reported in the aforementioned studies were kept in the analysis. The reason is that the above process gave a negative \gls{vebe} value for the rest of the geometries. A negative \gls{vebe} is unphysical since it implies that the electron would escape from the cluster but is kept there due to the finite basis set \cite{Herbert2005}.

\begin{table} [!htb]
    \centering
    \begin{tabular}{lc}
        Property & Convergence criterion \\
        \hline
        Self-Consistent Field (SCF) energy & $10^{-8}$ $E_\mathrm{h}$ \\
        Total energy & $5\times 10^{-6}$ $E_\mathrm{h}$ \\
        Maximum gradient & $3 \times 10^{-4}$ $E_\mathrm{h}$/$a_0$ \\
        Root mean square gradient & $1 \times 10^{-4}$ $E_\mathrm{h}$/$a_0$ \\
        Maximum displacement & $4 \times 10^{-3}$ $a_0$ \\
        Root mean square displacement & $2 \times 10^{-3}$ $a_0$ \\
        \hline
    \end{tabular}
    \caption{Convergence criteria in atomic units for the geometry optimization \gls{qm} computations.}
    \label{tab:go_conv_crit}
\end{table}

For the spin-unrestricted \gls{aimd} simulations presented in \cref{fig:spindens}, a timestep of 1 fs is used. The total simulated time is 500 fs. The initial velocities were randomly sampled from a \gls{mb} distribution that matches an initial temperature of 10 K. The temperature of the system was 100 K, regulated with the Berendsen thermostat, and a damping constant of 10 fs. The charge of the system was $-2$ $e$ and the spin multiplicity is equal to 3.

\textcolor{black}{Existing analytical formulations of Electron Field Emission (EFE) (e.g. Refs. \citenum{Stefanovic2006,Vaverka2014,Kyritsakis2015,Holgate2017a,Holgate2018}) are based on the free-electron model, which is only appropriate for metallic surfaces. Below, we show that these formulations cannot be applied to the grains investigated in this study because of the free-electron approximation. Our water-based grains behave as insulators at nano timescales because the solvated electrons are not free to move, but instead are strongly bound to the water cluster. Even though water-based grains can achieve the charge distribution of a conductor at macro timescales, as shown in \cref{sec:solv_cond}, they achieve this due to the slow diffusion of solvated electrons. Therefore, the EFE formulation is not applicable. We demonstrate this by applying an EFE formulation for metallic spheres\cite{Holgate2018} to the geometry of our \gls{aimd} simulations shown in \cref{fig:n24_21_24_spindens}.} 

\textcolor{black}{Recently, Holgate and Coppins \cite{Holgate2018} derived a Fowler-Nordheim (FN)-type\cite{Fowler1928} equation for an electrically isolated metallic sphere. The FN equation is derived for zero temperature, but at ordinary temperatures, the predictions of the FN equation do not vary significantly with temperature \cite{Fowler1928}. Therefore, the FN equation is appropriate for low-temperature metallic grains.}

\textcolor{black}{The FN type emitted current density from a metallic spherical surface is given by}\cite{Holgate2018}
\begin{equation}
    J_\mathrm{FN} (\mathcal{E}) = \frac{a_{\mathrm{FN}}}{\left[ {t^{*}_i}  (y_{\mathrm{FN}}, A_{\mathrm{FN}})
  \right]^2}  \frac{\mathcal{E}^2}{\mathrm{VEBE}} \exp \left[ - b_{\mathrm{FN}} 
  \frac{\mathrm{VEBE}^{3/2}}{\mathcal{E}} v_i^{*} (y_{\mathrm{FN}}, A_{\mathrm{FN}}) \right],
  \label{sieq:J_FN}
\end{equation}
\textcolor{black}{where the barrier form correction factor is given by}
\begin{equation}
    v_i^{*} (y_{\mathrm{FN}}, A_{\mathrm{FN}}) = \frac{3}{2} A_{\mathrm{FN}} y_{\mathrm{FN}} \int_{\rho_1}^{\rho_2} \left[ 1 -
  \frac{y_{\mathrm{FN}}}{2 A_{\mathrm{FN}} \rho^2 (\rho^2 - 1)} + \frac{y_{\mathrm{FN}}}{A_{\mathrm{FN}}} \left( 1 - \frac{1}{\rho}
  \right) - A_{\mathrm{FN}} y_{\mathrm{FN}} \left( 1 - \frac{1}{\rho} \right) \right]^{1 / 2} d \rho
\end{equation}
\textcolor{black}{and the second barrier form correction factor is}
\begin{equation}
  t_i^{*} (y_{\mathrm{FN}}, A_{\mathrm{FN}}) = v_i^{*} (y_{\mathrm{FN}}, A_{\mathrm{FN}}) - \frac{2 y_{\mathrm{FN}}}{3}
  \frac{\partial v_i^{*} (y_{\mathrm{FN}}, A_{\mathrm{FN}})}{\partial y_{\mathrm{FN}}}.
\end{equation}
\textcolor{black}{In the above equations, $\rho = r/r_d$ is the dimensionless $r$ coordinate of the sphere in spherical coordinates. Also, the $y_{\mathrm{FN}}$ and $A_{\mathrm{FN}}$ parameters are given by}
\begin{equation}
    y_{\mathrm{FN}} = \frac{c_{\mathrm{Sc}} \sqrt{\mathcal{E}}}{\mathrm{VEBE}}
\end{equation}
and 
\begin{equation}
    A_{\mathrm{FN}} = \frac{e r_d  \sqrt{\mathcal{E}}}{c_{\mathrm{Sc}}}.
\end{equation}
\textcolor{black}{The integration limits $\rho_1$ and $\rho_2$ correspond to the roots of the equation} \cite{Holgate2017a}
\begin{equation}
  1 - \frac{y_{\mathrm{FN}}}{2 A_{\mathrm{FN}} \rho^2 (\rho^2 - 1)} + \frac{y_{\mathrm{FN}}}{A_{\mathrm{FN}}} \left( 1 -
  \frac{1}{\rho} \right) - A_{\mathrm{FN}} y_{\mathrm{FN}} \left( 1 - \frac{1}{\rho} \right) = 0.
\end{equation}
\textcolor{black}{The above constants are $c_{\mathrm{Sc}} = \sqrt{e^3/(4 \pi \epsilon_0)} \approx 1.199985$ eV(V/nm)$^{-1/2}$, $z_\mathrm{S} = 4 \pi e m_e/h^3 \approx 1.618311 \times 10^{14}$ A m$^{-2}$ eV$^{-2}$, $g_e = \sqrt{8 m}/\hbar \approx 10.24633$ eV$^{-1/2}$ nm$^{- 1}$, $a_{\mathrm{FN}} = z_\mathrm{S} [2/(3 b_{\mathrm{FN}}) ]^2 \approx 1.541434$ \textmu A eV V$^{-2}$ and $b_{\mathrm{FN}} = 2 g_e/(3 e) \approx 6.830890$ (eV)$^{-3/2}$V nm$^{-1}$.}

\textcolor{black}{As noted by Holgate and Coppins \cite{Holgate2018}, $J_{FN}$ of \cref{sieq:J_FN} may not give accurate results at radii of curvature smaller than 2 nm, because quantum confinement effects are neglected. Here, we apply \cref{sieq:J_FN} to the biggest AIMD structure of our study, shown in \cref{fig:n24_21_24_spindens}, even though it does not satisfy this limit. We assume that the grain in \cref{fig:n24_21_24_spindens} is spherical with $r_d\approx 10$ {\AA} and $Q=-2$~$e$. The VEBE of each solvated electron is 0.78 eV. Based on these values, $J_{\mathrm{FN}}\approx10^{11}$ A/m$^2$, and the corresponding ion and electron currents predicted by the OML theory are respectively $I_i \approx 10^{-16}$ A/m$^2$ and $I_e\approx -10^{-15}$ A/m$^2$. The quantity $J_{FN}A$ gives an estimate of the number of emitted electrons per unit time and is equal to $J_{FN}A\approx 10^4$ $e$/ns.} \textcolor{black}{The FN equation predicts that the solvated electrons should immediately escape the water molecule cluster, which contradicts our AIMD simulations.} 

\textcolor{black}{Consider also the following case, which falls within the validity range of \cref{sieq:J_FN} i.e., a sphere with $r_d\ge2$ nm \cite{Holgate2018}. Suppose we have a spherical grain charged with $Q=-12$ $e$, where all electrons reside on its surface. Electrons will be arranged in the configuration with the minimum potential energy. This is a Thomson problem \cite{Thomson1904} and has an exact solution for 12 electrons. Specifically, the electrons will reside on the vertices of a regular icosahedron \cite{Andreev1996}. The radius of the circumsphere of a regular icosahedron with edge length $l$ is $r_d=l(\phi_g^2+1)^{1/2}/2$, where $\phi_g=(1+\sqrt{5})/2$. Assuming an edge distance 3 times larger than the electron-electron distance in our AIMD simulation of \cref{fig:n24_21_24_spindens}, i.e., $l=3d_{e-e}\approx60$ \AA, we get $r_d\approx57$ {\AA}. Based on those values, our AIMD simulations indicate negligible electron-electron repulsions and thus no EFE. However, \cref{sieq:J_FN} predicts that $J_{FN}A\approx 10^4$ $e$/ns, which suggests strong EFE.}

\textcolor{black}{Consequently, it is evident that the metal-based EFE formulations significantly overestimate electron-electron repulsions of water grains and thus cannot be applied to our study.}

\section{Surface tension of neutral and charged nanoclusters}
\label{sec:si_surf_tens}

\begin{figure}[!htb]
    \centering
    \includegraphics[width=0.5\linewidth]{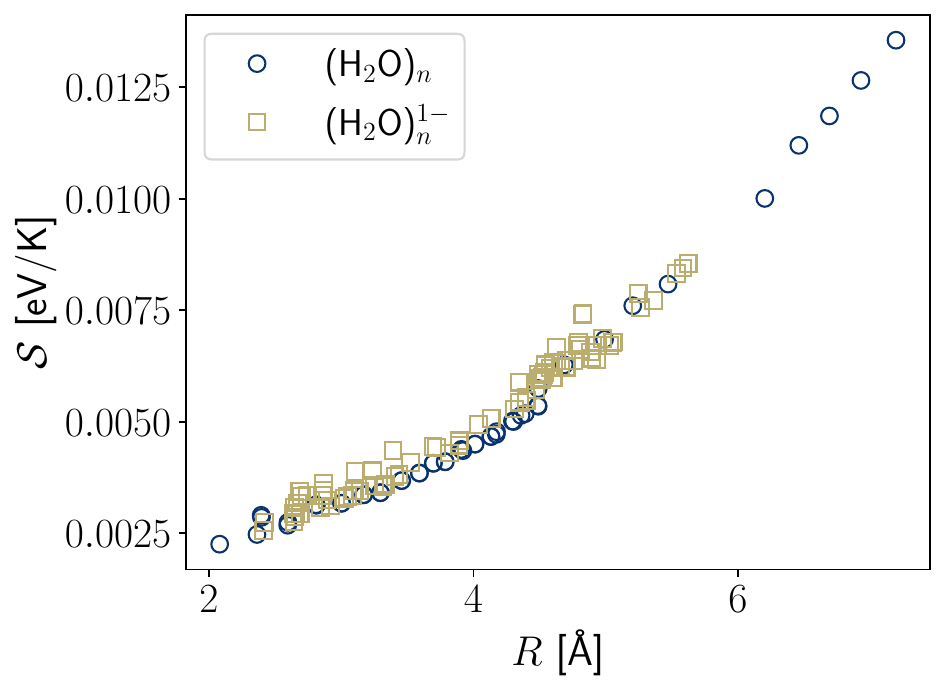}%
    \includegraphics[width=0.48\linewidth]{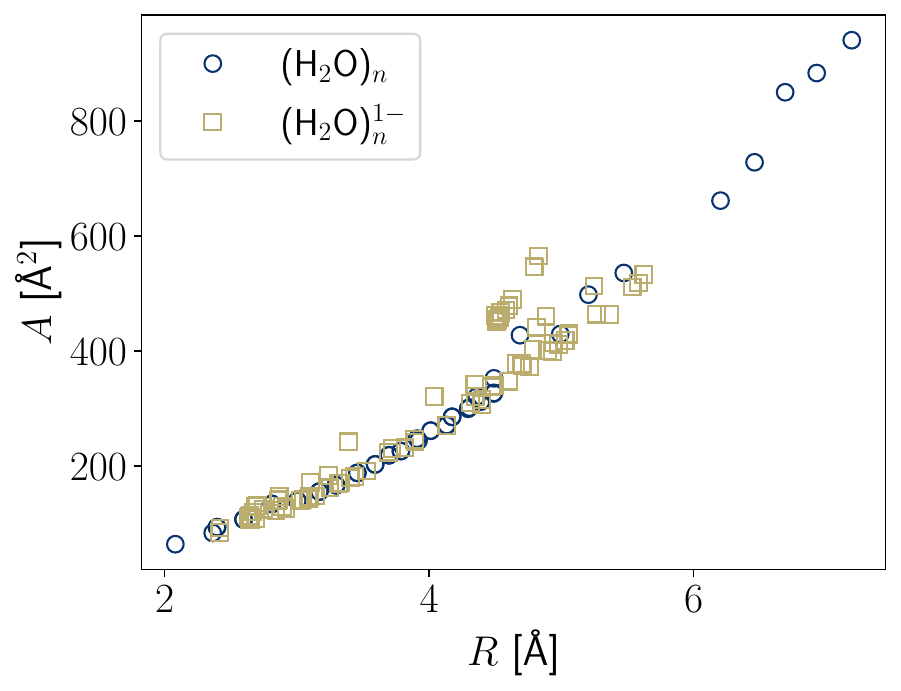}
    \caption{Entropy (left) and area (right) of the neutral and charged water clusters shown in \cref{fig:surface_tension} of the manuscript.}
    \label{fig:entropy_area}
\end{figure}

The surface tension is computed for neutral and charged water clusters. A total of 35 geometries of neutral clusters were taken from Refs. \citenum{wales1998,rakshit2019}. In addition, 11 geometries ($n=3$, 18, 21, 24, 28, 31, 44, 49, 55, 60, 66) were generated from a bulk structure similar to the process detailed in \cref{sec:si_solv_cond}. In addition, 96 geometries of negative clusters were taken from Refs. \citenum{Zhan2003,Williams2008,Herbert2005,Kumar2015} that had a positive \gls{vebe} at the \gls{dft}/B3LYP-D3/6-311++G** level of theory. Three geometries of charged clusters, with $n=8$, 16, and 28, were generated from a bulk structure. \Cref{fig:entropy_area} shows the entropy and area of the various neutral and charged clusters investigated in this study.

%